# The dynamics of dust particles during the laser-induced plasma process: Self-similar expansion in a liquid

**M. S. Afify[1] and S. Salem[2]**

[1]Department of Physics, Faculty of Science, Benha University, Benha 13518, Egypt

[2]Basic and Applied Science Department, College of Engineering and Technology, Arab Academy for Science and Technology (AAST), Port Said, Egypt

Corresponding author: mahmoud.afify@fsc.bu.edu.eg

## Abstract

The role of liquid ions in the dusty plasma produced by laser ablation in a liquid environment is still unclear. For that purpose, we utilized the self-similar approach to investigate the effect of liquid ions on plasma expansion front. The fluid equations have been used for dust particles and ion species in addition to superthermal electron distribution. We treat the expansion process in the absence and existence of liquid ions. The impact of the dust-to-ion initial number density ratio, superthermality factor kappa, dust-to-electrons initial temperature ratio, and dust-ions atomic number on the expansion front has been addressed. Our findings show that in the single ion case, changing the superthermality of energetic electrons has the most significant influence on the dust ion expansion. Moreover, we observed that the mass of the liquid ions has the greatest effect on the dynamics of the dust ions in the double ion case.

## I. Introduction

Research on the interaction between laser and plasmas started in the 1970s, propelled by the induction of nuclear fusion by means of laser [1]. Nanosecond lasers have been utilized for laser produced plasma (LPP) [2], however, the interaction between laser and targets have different challenges such as the weak absorption of the laser energy [3] and changing the characteristics of the targets because of heating of the objective material [4]. This is a result of the long-duration nanosecond laser signal contrasted with the relaxation time among lattice and electrons [5].

However, in the 1990s, super high pinnacle power and super short pulse lasers were acknowledged with the innovation of the chirped pulse amplification (CPA) strategy [6] combined with the creation of the Kerr lens mode-locking procedure [7] joined with the invention of the titanium: sapphire (Ti: sapphire) laser medium [8]. This technology leads to reduce the laser pulse duration compared to the relaxation time of electrons and lattice, hence reducing the heating of the target.

Laser-matter interaction has the potential for significant applications in scientific and industrial fields [9]. The generation of protons from the high intensity laser-matter interaction opens the door for new area of application of radiography [10, 11]. The extreme ultraviolet (EUV) plasma emission is an interesting application of high intensity laser-matter interaction where the laser beams are used to implode a small solid target to reach a density and temperature comparable of solid material and Fermi temperature. Moreover, the production of EUV lithography is a promising technique for high-volume semiconductor manufactures such as computer chips [12, 13, 14]. Achieving the state of warm dense matter (WDM) to study the equation of state (EOS), providing the ignition energy of fusion targets, and the cancer therapy utilizing plasmas are also an important application of the interaction between high intensity laser beam and high dense matter [see, e.g., Refs. 15 and 16 and references therein].

Recently, the laser-matter interaction underwater had an important consideration of research as it can be utilized for nanocrystals synthesis, biomedicine, and sensing [17]. When a high intensity laser beam is incident onto a thin target in a liquid, it causes the production of a hot plasma owing to the coupling of the focused laser energy with the target prompts quick heating to the surface. The liquid environment confines the expansion rate of the resulting plasma causing the nucleation of the dust particles. Moreover, the plasma density reaches a high value of more than $10^{25}$ $m^{-3}$ [18]. One more resulting impact of the confined high warm plasma is that a limited quantity of the encompassing fluid is disintegrated to shape a cavitation bubble between the plasma head and the fluid at around 1 μs. The bubble grows to the most extreme size, then, at that point, it therapists and implodes later around 300 μs. Although the formation of the cavitation bubble leads to accumulating the nanoparticles into its surface, the arrangement of nanoparticles occurs some time before, during the plasma-stage development. Interestingly, the shrinking of the bubble also is responsible for the compact of the nanoparticles [19].

Novel articles explored the dynamics of dust particles in liquid. The parameters that may lead to the production of quantum dots in liquid from CdS and ZnSe semiconductors was reported in Ref. [20]. Haustrup et al [21] used the experimental techniques to investigate the dynamics of different types of nano particles such as gold particles during the ablation of nano and femto laser beam in liquid. They concluded that the liquid environment controls the radius of the dust particles. Further, the mechanism behind the formation of the dust particle during the laser ablation of different targets had been examined in Refs. [22-26].

Interestingly, the process of laser ablation in liquid leads to the formation of a compressed high thermal dense plasma that is mixed with the plasma which is made by the fluid decay [27]. As a result, the dust particles are formed due to the development and nucleation of the plasma mixture. Therefore, our aim is to investigate the role of the secondary ions; i.e. liquid ions, beside the background charges during the plume expansion of the silver target according to Fig. 1. We employed the self-similar technique to investigate the dynamics of dust particles and the plasma density and velocity. This article is arranged as follows: The proposed model and scaling are adopted in Sec. II. The expansion of background plasma and the case of plasma mixture is examined Sec. III. Finally, the results are summarized in Sec. IV.

## II. Problem Formulation

The mechanism behind the plasma expansion during laser ablation in different mediums can be divided into two stages. The first phase results from the existence of self-electric fields. This is because the electrons move faster than the heavy ions giving rise to the creation of charge separation which is responsible for the acceleration of ions to maintain the condition of quasi-neutrality. The deposition of energy via the laser pulse is responsible for the second phase owing to the production of the electrostatic double layer. The latter results from the gradient of the thermal pressure; i.e. the gradient of the electron temperature, where electrons from the thermal plasma move to the cold regions and vice versa. Here, we utilized the self-similar method to examine the dynamics of the plasma during the plume expansion considering the expansion is homogeneous. In fact, the theory of laser-matter interaction is very complicated since the plasma is usually described by nonlinear partial differential equations. The Particle-In-Cell simulation is well familiar to study the expansion of laser ablation. However, targets with electron density equals to the cutoff density, $\left(n_c = \frac{m_e \omega^2}{4\pi e^2} = 1.1 \times 10^{21} cm^{-3} \left[\frac{\lambda}{1\mu m}\right]^{-2}\right)$, make the task of supercomputers is very hard. This is because a problem with spatial scales $\sim \frac{c}{\omega_p}$ and temporal scales $\sim \omega_p^{-1}$ will give rise to thousands of grid points in each spatial direction and thousands of time steps. Fortunately, the numerical studies reported that the state of the growing plasma front is coincidence with the overall highlights of the self-similar arrangement [29, 30]. Therefore, although the self-similar technique can't clarify the total dynamics of plasma development into fluid, yet it assists to catch the fundamental elements of the issue.

We assumed a collisionless plasma which is consists of negative dust particles, two positive ions, and superthermal electrons occupied the space of $x < 0$ at $t = 0$, while the liquid is considered to occupy the space of $x > 0$. The dynamics of the dust particles at $t > 0$ can be obtained through the following multi-fluid equations:

$$\frac{\partial n_d}{\partial t} + \frac{\partial}{\partial x}(n_d u_d) = 0, \qquad (1)$$

$$m_d n_d \left(\frac{\partial u_d}{\partial t} + u_d \frac{\partial u_d}{\partial x}\right) = z_d e n_d \frac{\partial \varphi}{\partial x} - \frac{\partial P_d}{\partial x}, \qquad (2)$$

where the index d refers to the dust particles. The quantities $m_d$, $n_d$, and $u_d$ are the dust mass, number density, and velocity, respectively. The symbol $\varphi$ is the electrostatic potential and $P_d$ is the dust pressure gradient. $z_d$ refers to the charge number. Also, for the positive species, we have:

$$\frac{\partial n_j}{\partial t}+\frac{\partial}{\partial x}\left(n_j u_j\right) \quad = 0, \qquad (3)$$

$$m_j n_j\left(\frac{\partial u_j}{\partial t}+u_j\frac{\partial u_j}{\partial x}\right) \quad = -e n_j\frac{\partial \varphi}{\partial x}-\frac{\partial P_j}{\partial x}, \qquad (4)$$

where the subscription $j = 1{,}2$ denote the two ion species 1 and 2 and $n_j$ and $u_j$ are the ion-number density and velocity, respectively. $P_j$ is the ion gradient pressure, $m_j$ refers to the ion mass. This system is enclosed by the following Poisson's equation:

$$\frac{\partial^2 \varphi}{\partial x^2}=\frac{e}{\varepsilon_0}\left(n_1+n_2-z_d n_d-n_e\right), \qquad (5)$$

The density of electrons can be described by the following "kappa" distribution [31]:

$$n_e = n_{e0}\left(1-\frac{e\Phi}{T_e\left(\kappa-\frac{3}{2}\right)}\right)^{\left(-\kappa+\frac{1}{2}\right)}, \qquad (6)$$

where $\kappa$ refers to the spectral index which is a scale of the superthermality, $T_e$ is the electron temperature, and $e$ denotes the electron charge.

Interestingly, Vasyliunas [32] was the first to propose the Kappa distribution to address the dynamics of superthermal particles in the night area of the magnetosphere. This kind of distribution function has been utilized for displaying non-Maxwellian foundations in different astrophysical and bounded plasma [33, 34, 35]. The Kappa conveyance work relies upon a genuine boundary ($\kappa$), which estimates the extent of the superthermal tail in the distribution function, addressing the abundance of exceptionally enthusiastic particles. Increasing the values of $\kappa$ expands, leads to shrinking the superthermal tail, and the Maxwellian dispersion can be recuperated at $\kappa \to \infty$.

Here, we assume the plasma component should achieve the following neutrality condition

$$z_d n_d + n_e = n_1 + n_2. \tag{7}$$

This is because the characteristic of the quasi-neutrality remains useful during the expansion process as the Debye length is smaller than the scale length of the density gradient. Now, it is convenient to express the last equations in the normalized form for the analytical analysis according to the following table:

Table 1 The normalization of variables.

| Quantity | Symbol | Normalized by | Formula |
|---|---|---|---|
| Time | $t$ | Inverse of plasma frequency | $\omega_{p1}^{-1} = \left(\frac{\varepsilon_0 m_1}{e^2 n_{10}}\right)^{-\frac{1}{2}}$ |
| Space | $x$ | Debye radius | $\lambda_{D1} = \left(\frac{k_B T_e \varepsilon_0}{e^2 n_{10}}\right)^{\frac{1}{2}}$ |
| Velocity | $u$ | Ion acoustic speed | $C_s = \left(\frac{k_B T_e}{m_1}\right)^{\frac{1}{2}}$ |
| Potential | $\varphi$ | Thermal potential | $\frac{k_B T_e}{e}$ |

Applying this normalization scheme, we obtain the following dimensionless normalized differential equations.

The form of the dust fluid equations is:

$$\frac{\partial N_\mathrm{d}}{\partial t} + \frac{\partial}{\partial x} N_\mathrm{d} V_\mathrm{d} = 0, \tag{8}$$

$$\left(\frac{\partial}{\partial t} + V_\mathrm{d}\frac{\partial}{\partial x}\right) V_\mathrm{d} - Z_\mathrm{d} Q_\mathrm{d}\frac{\partial \phi}{\partial x} + 3\sigma_\mathrm{d} Q_\mathrm{d} N_\mathrm{d}\frac{\partial N_\mathrm{d}}{\partial x} = 0, \tag{9}$$

For the first positive ions, we have:

$$\frac{\partial N_1}{\partial t} + \frac{\partial}{\partial x} N_1 V_1 = 0, \tag{10}$$

$$\left(\frac{\partial}{\partial t}+V_1\frac{\partial}{\partial x}\right)V_1+\frac{\partial\phi}{\partial x}+3\sigma_1N_1\frac{\partial N_1}{\partial x}=0, \tag{11}$$

Also, the second positive ions can be expressed by the following equations:

$$\frac{\partial N_2}{\partial t}+\frac{\partial}{\partial x}N_2V_2=0, \tag{12}$$

$$\left(\frac{\partial}{\partial t}+V_2\frac{\partial}{\partial x}\right)V_2+Q_{\mathrm{i}}\frac{\partial\phi}{\partial x}+3\sigma_2Q_{\mathrm{i}}N_2\frac{\partial N_2}{\partial x}=0, \tag{13}$$

Finally, the dimensionless form of the superthermal background electrons is:

$$N_e=\left(1-\frac{\phi}{\left(\kappa-\frac{3}{2}\right)}\right)^{-\kappa+\frac{1}{2}}, \tag{14}$$

where we considered a 1-D adiabatic pressure $P_{d,1,2}\equiv k_BT_{d,1,2}n_{d,1,2}^{\Gamma}$. For $N=1$, the value of the constant $\Gamma=\frac{2+N}{N}$ is 3 where $N$ refers to the possible degrees of freedom and $k_B$ refers to the Boltzmann constant. The density, $N_{\mathrm{d},1,2,e}$, is normalized wih respect to the unperturbed density $n_{\mathrm{d}0,10,20,e0}$. The symbols $\sigma_{\mathrm{d}}$, $\sigma_1$, $\sigma_2$, Q, and $Q_{\mathrm{i}}$ refers to the ratios $\frac{T_{\mathrm{d}}}{T_e}$, $\frac{T_1}{T_e}$, $\frac{T_2}{T_e}$, $\frac{m_1}{m_d}$, and $\frac{m_1}{m_2}$, respectively.

## III. The Scheme of the Self-Similar Technique

In this section, we introduce the variable $\xi=x/c_st$ [31] into the Eqns. (8-14) to convert them from partial differential equations to ordinary differential equations. Moreover, we utilized this variable in both the neutrality condition which is given by Eq. (7). Therefore, the full scheme of the self-similar solution can be given by the following set of ordinary normalized dimensionless differential equations:

$$(V_{\mathrm{d}}-\xi)\frac{\mathrm{d}N_{\mathrm{d}}}{\mathrm{d}\xi}+N_{\mathrm{d}}\frac{\mathrm{d}}{\mathrm{d}\xi}V_{\mathrm{d}}=0, \tag{15}$$

$$(V_{\mathrm{d}}-\xi)\frac{\mathrm{d}V_{\mathrm{d}}}{\mathrm{d}\xi}-Z_{\mathrm{d}}Q_{\mathrm{d}}\frac{\mathrm{d}\phi}{\mathrm{d}\xi}+3\sigma_{\mathrm{d}}Q_{\mathrm{d}}N_{\mathrm{d}}\frac{\mathrm{d}N_{\mathrm{d}}}{\mathrm{d}\xi}=0, \tag{16}$$

$$(V_1-\xi)\frac{\mathrm{d}N_1}{\mathrm{d}\xi}+N_1\frac{\mathrm{d}}{\mathrm{d}\xi}V_1=0, \tag{17}$$

$$(V_1-\xi)\frac{\mathrm{d}V_1}{\mathrm{d}\xi}+\frac{\mathrm{d}\phi}{\mathrm{d}\xi}+3\sigma_1 N_1\frac{\mathrm{d}N_1}{\mathrm{d}\xi}=0, \qquad (18)$$

$$(V_2-\xi)\frac{\mathrm{d}N_2}{\mathrm{d}\xi}+N_2\frac{\mathrm{d}}{\mathrm{d}\xi}V_2=0, \qquad (19)$$

$$(V_2-\xi)\frac{\mathrm{d}V_2}{\mathrm{d}\xi}+Q_{\mathrm{i}}\frac{\mathrm{d}\phi}{\mathrm{d}\xi}+3\sigma_2 Q_{\mathrm{i}} N_2\frac{\mathrm{d}N_2}{\mathrm{d}\xi}=0, \qquad (20)$$

$$\frac{1}{N_e}\frac{dN_e}{d\xi}-\frac{\left(\kappa-\frac{1}{2}\right)}{\left(\kappa-\frac{3}{2}-\Phi\right)}\frac{d\Phi}{d\xi}=0. \qquad (21)$$

$$\frac{\mathrm{d}N_1}{\mathrm{d}\xi}+\alpha\frac{\mathrm{d}N_2}{\mathrm{d}\xi}-\beta\frac{\mathrm{d}N_{\mathrm{d}}}{\mathrm{d}\xi}-\gamma\frac{\mathrm{d}N_{\mathrm{e}}}{\mathrm{d}\xi}=0, \qquad (22)$$

where $\alpha=n_{20}/n_{10}, \beta=n_{\mathrm{d}0}/n_{10},$ and $\gamma=n_{e0}/n_{10}$. The last arrangement of equations will shape the functioning tool compartment for the examination that follows. Prior to continuing with our investigation of the dynamics of two positive ions, we will consider the single positive ion restricting case, as it gets from the above model to acquire an important understanding of the model and its limits.

### *A. Single positive ion case*

For the single ions, we considered $N_2=V_2=0$ where Eqs. (15 - 22) reduce to

$$(V_{\mathrm{d}}-\xi)\frac{\mathrm{d}N_{\mathrm{d}}}{\mathrm{d}\xi}+N_{\mathrm{d}}\frac{\mathrm{d}}{\mathrm{d}\xi}V_{\mathrm{d}}=0,$$

$$(V_{\mathrm{d}}-\xi)\frac{\mathrm{d}V_{\mathrm{d}}}{\mathrm{d}\xi}-Z_{\mathrm{d}}Q_{\mathrm{d}}\frac{\mathrm{d}\phi}{\mathrm{d}\xi}+3\sigma_{\mathrm{d}}Q_{\mathrm{d}}N_{\mathrm{d}}\frac{\mathrm{d}N_{\mathrm{d}}}{\mathrm{d}\xi}=0,$$

$$(V_1-\xi)\frac{\mathrm{d}N_1}{\mathrm{d}\xi}+N_1\frac{\mathrm{d}}{\mathrm{d}\xi}V_1=0,$$

$$(V_1-\xi)\frac{\mathrm{d}V_1}{\mathrm{d}\xi}+\frac{\mathrm{d}\phi}{\mathrm{d}\xi}+3\sigma_1 N_1\frac{\mathrm{d}N_1}{\mathrm{d}\xi}=0,$$

$$\frac{1}{N_e}\frac{dN_e}{d\xi}-\frac{\left(\kappa-\frac{1}{2}\right)}{\left(\kappa-\frac{3}{2}-\Phi\right)}\frac{d\Phi}{d\xi}=0.$$

$$\frac{\mathrm{d}N_1}{\mathrm{d}\xi} + \alpha\frac{\mathrm{d}N_2}{\mathrm{d}\xi} - \beta\frac{\mathrm{d}N_\mathrm{d}}{\mathrm{d}\xi} - \gamma\frac{\mathrm{d}N_\mathrm{e}}{\mathrm{d}\xi} = 0,$$

Although the self-similar technique is powerful in paving the way to solve the system of the partial differential equations, it also diminishes some details of plasma system such as the exact variables of length and time and replace them with the less informative self-similar variable ($\xi$). We used numerical code and implemented the "NDsolve" command included in Wolfram Mathematica package we then used the following initial conditions $n_1[0] = 0.1$, $n_d[0] = 0.1$, $n_e[0] = 0.2$, $u_1[0] = 1.01$, $u_d[0] = 1.01$, and $\phi[0] = 0$ to be able to solve the set of ordinary differential equations with the independent variables.

We investigated the effects of four different plasma variables which are (dust-to-ion initial number density ratio, super-thermality factor kappa, dust-to-electrons initial temperature ratio, and lastly dust ions atomic number) on the negative dust density and velocity variations as well as the variation of the electric potential generated through the plasma.

According to our investigations, the plasma expansion is slightly affected by the dust to ion initial number density ratio variation. In Fig. 2(a) we can observe that the depletion of dust ion density slightly variates versus the self-similar variable ($\xi$). While in Fig. 2(b) the dust ion velocity fairly changes with ($\xi$). The electric potential changes directly with increasing the dust-to-ion initial density versus ($\xi$) as depicted in Fig. 2(c).

Moreover, the plasma expansion is relatively heavily affected by the superthermality factor ($\kappa$) as we can clearly see in Fig. 3. The reason behind this could be due to the high energetic superthermal electrons that can generate an intense electric field which affects the dust ion in such way. In Fig. 3(a) the dust density ($n_d$) changes versus ($\xi$) inversely with increasing the superthermality factor while the dust ion freedom, i.e. velocity ($u_d$) changes directly with increasing the superthermality factor as shown in Fig. 3(b). Figure 3(c) shows that the electric potential increases with increasing the superthermal ($\kappa$) factor versus ($\xi$).

In Figs. 4 and 5 there is no change in plasma expansion variation corresponding to the change of dust ion atomic number ($Z_d$) and dust-to-electron temperature ratio ($\sigma_d$) respectively.

### *B. Two positive ions case*

Here, we considered the impact of the second ions where system of Eqs. (15 - 22) can be now fully considered and solved numerically to obtain a glimpse of the second ions' characteristics influences on the dust ions' dynamics. In order to obtain the results we have used the numerical code and the implemented "NDsolve" command included in Wolfram Mathematica package software, we then used the following initial conditions $n_1[0] = 0.1$, $n_2[0] = 0.1$, $n_d[0] = 0.1$, $n_e[0] = 0.2$, $u_1[0] = 1.01$, $u_2[0] = 1.01$, $u_d[0] = 1.01$, and $\phi[0] = 0$ to be able to solve the set of ordinary differential equations with the independent variables.

We investigated the effect of four different plasma variables of the second positive ions and their effect on the dust ions density and velocity and also on the overall generated electric potential through the plasma. The plasma variables are (second-to-first ion initial number density ratio, second ion to electron temperature density, second ions mass, and lastly second ions atomic number). According to our investigations, the plasma expansion of dust ions is affected by the second to the first ion density ratio variation as one can see in Fig. 6, that the larger the ratio (high second ion number density) the less expandable the plasma. In Fig. 6(a) we can see that the depletion of dust ion density variates versus the self-similar variable ($\xi$). While in Fig. 6(b) the dust ion velocity changes with ($\xi$). And in Fig. 6(c) the electric potential changes directly with increasing the second to the first ion density ratio versus ($\xi$).

Also, the plasma expansion is relatively affected by the first ion to second ion mass ratio $Q_2$ as we can clearly see in figure 7 in such a way that the lighter the second ion could get the depletion in the dust ion expansion could be increased. In Fig. 7(a) the dust density ($n_d$) changes versus ($\xi$) directly with increasing the mass ratio $Q_2$ while the dust ion freedom, i.e. velocity ($u_d$) changes directly with increasing the mass ratio $Q_2$ factor as shown in Fig. 7(b). And in Fig. 7(c) the electric potential increases with increasing the mass ratio $Q_2$versus ($\xi$).

In Fig. 8 we can depict that the plasma expansion of dust ions is slightly affected by the second ion to electron temperature ratio variation, that the larger the ratio (i.e. the higher temperature the second ion could get) the more expandable and the faster the plasma could reach. Figure 8(a) shows that the depletion of dust ion density slightly varies versus the self-similar variable ($\xi$). While in Fig. 8(b) the dust ion velocity fairly changes with ($\xi$). And in Fig. 8(c) the electric potential changes directly with increasing the dust-to-ion initial density versus ($\xi$). Figure 9 presents that the plasma expansion of dust ions is investigated against the change in the second ion atomic number $Z_2$ variation. It's observed that the second ion atomic number has no effect for the low values of $Z_2$ till it reaches some certain value that is comparable to the first ion atomic number. Figure 9(a) showed that the depletion of dust ion density variates versus the self-similar variable ($\xi$). While in Fig. 9(b) the dust ion velocity changes with ($\xi$). And in Fig. 9(c) the electric potential changes directly with increasing the second to the first ion density ratio versus ($\xi$).

## IV. Conclusion

In this work, we investigated the role of the secondary ions during the expansion process in the liquid environment. At the first stage of the expansion process, we considered three components, which are the fluid positive ions, the fluid negative dust particles, and the superthermal electrons that are presented by the kappa distribution. Then, we included the impact of the secondary positive ions along with the three components in the second stage. We investigated the roles of the dust-to-ion initial number density ratio, the superthermality factor kappa, the dust-to-electrons initial temperature ratio, and lastly, the dust-ions atomic number on the plasma expansion front. The results showed that changing the superthermality of energetic electrons has the most significant influence on the dust ion expansion in the absence of the secondary ions. While the mass of the second ion has the greatest effect on the dust ion dynamics. Finally, we stressed that the self-similar approach we used in our research has its own set of limitations and potential flaws. This is because it provides a specific solution to the fluid equations and therefore does not provide a thorough description of the complete process. However, the self-similar solution is worthwhile and valuable for the straightforward and analytically feasible formulation of plasma expansion.

## Figure captions

Fig. 1. A sketch of the first and second stages during the laser-silver target interaction in liquid: The red circles refer to the induced plasma and dust particles and the orange circles refer to the mixed plasma, i.e. the induced plasma, dust particles, and liquid ions.

Fig. 2. (a) Normalized dust density $n_d$, (b) normalized dust velocity $u_d$, and (c) normalized electric potential $\phi$ versus the self-similar variable $\xi$, For different values of dust-to-ion initial number density, where $\beta = 0.5$ (solid black lines), $\beta = 0.6$ (red dashed lines), $\beta = 0.7$ (blue dotted lines), and $\beta = 79$ (orange dot-dashed lines). Here, $Q_d = 0.000611$, $\sigma_i = \sigma_d = 0.1$, $\kappa = 2$, $Z_i = 47$, and $Z_d = 21$.

Fig. 3. (a) Normalized dust density $n_d$, (b) normalized dust velocity $u_d$, and (c) normalized electric potential $\phi$ versus the self-similar variable $\xi$, For different values of superthermality parameter $\kappa$, where $\kappa = 2$ (solid black lines), $\kappa = 5$ (red dashed lines), $\kappa = 8$ (blue dotted lines), and $\kappa = 10$ (orange dot-dashed lines). Here, $Q_d = 0.000611$, $\sigma_i = \sigma_d = 0.1$, $\beta = 0.5$, $Z_i = 47$, and $Z_d = 21$.

Fig. 4. Normalized positive ion density $n_d$ versus the self-similar variable $\xi$, Here, $Q_d = 0.000611$, $\sigma_i = 0.1$, $\beta = 0.5$, $Z_i = 47$, $\kappa = 2$, and $Z_d = 21$.

Fig. 5. Normalized positive ion density $n_d$ versus the self-similar variable $\xi$, Here, $Q_d = 0.000611$, $\sigma_i = \sigma_d = 0.1$, $\beta = 0.5$, $Z_i = 47$, and $\kappa = 2$.

Fig. 6. (a) Normalized dust density $n_d$, (b) normalized dust velocity $u_d$, and (c) normalized electric potential $\phi$ versus the self-similar variable $\xi$, For different values of second ion initial number density ratio $\alpha$, where $\alpha = 0.5$ (solid black lines), $\alpha = 1$ (red dashed lines), $\alpha = 1.5$ (blue dotted lines), and $\alpha = 2$ (orange dot-dashed lines). Here, $Q_2 = 1$, $Q_d = 0.000611$, $\sigma_1 = \sigma_2 = \sigma_d = 0.1$, $\kappa = 2$, $Z_1 = Z_2 = 47$, $\beta = 0.5$, and $Z_d = 21$.

Fig. 7. (a) Normalized dust density $n_d$, (b) normalized dust velocity $u_d$, and (c) normalized electric potential $\phi$ versus the self-similar variable $\xi$, For different values of first ion to second ion mass ratio $Q_2$, where $Q_2 = 0.1$ (solid black lines), $Q_2 = 1$ (red dashed lines), $Q_2 = 2$ (blue dotted lines), and $Q_2 = 4$ (orange dot-dashed lines). Here, $\alpha = 0.5$ , $Q_d = 0.000611$, $\sigma_1 = \sigma_2 = \sigma_d = 0.1$, $\kappa = 2$, $Z_1 = Z_2 = 47$, $\beta = 0.5$, and $Z_d = 21$

Fig. 8. (a) Normalized dust density $n_d$, (b) normalized dust velocity $u_d$, and (c) normalized electric potential $\phi$ versus the self-similar variable $\xi$, For different values of second ion to electron temperature ratio $\sigma_2$, where $\sigma_2 = 0.001$ (solid black lines), $\sigma_2 = 0.01$ (red dashed lines), $\sigma_2 = 0.1$ (blue dotted lines), and $\sigma_2 = 0.2$ (orange dot-dashed lines). Here, $\alpha = 0.5$, $Q_2 = 1$, $Q_d = 0.000611$, $\sigma_1 = \sigma_d = 0.1$, $\kappa = 2$, $Z_1 = Z_2 = 47$, $\beta = 0.5$, and $Z_d = 21$

Fig. 9. (a) Normalized dust density $n_d$, (b) normalized dust velocity $u_d$, and (c) normalized electric potential $\phi$ versus the self-similar variable $\xi$, For different values of second ion atomic number $Z_2$, where $Z_2 = 30$ (solid black lines), $Z_2 = 41$ (red dashed lines), $Z_2 = 44$ (blue dotted lines), and $Z_2 = 45$ (orange dot-dashed lines). Here, $\alpha = 0.5$, $Q_2 = 1$, $Q_d = 0.000611$, $\sigma_1 = \sigma_d = 0.1$, $\kappa = 2$, $Z_1 = 47$, $\beta = 0.5$, and $Z_d = 21$

## Reference

[1] J. Nuckolls, L. Wood, A. Thiessen, and G. Zimmerman, Nature 239, 139-42 (1972).

[2] M. Elsied, P. K. Diwakar, M. Polek, and A. Hassanein, J. Appl. Phys. 120, 173104 (2016).

[3] X. Ni, K. K. Anoop, X. Wang, D. Paparo, S. Amoruso, and R. Bruzzese, Appl. Phys. A Mater. Sci. Process. 117, 111-115 (2014).

[4] B. N. Chichkov, C. Momma, S. Nolte, F. Alvensleben, and A. Tünnermann, Appl. Phys. A Mater. Sci. Process. 63, 109-115 (1996).

[5] E. G. Gamaly, A.V. Rode, B. Luther-Davies, and V. T. Tikhonchuk, Phys. Plasmas 9, 949-957 (2002).

[6] D. Strickland and G. Mourou, Opt. Commun. 55, 447-9 (1985).

[7] D. E. Spence, P. N. Kean, W. Sibbett, Opt. Lett. 16, 42-4 (1991).

[8] P. F. Moulton, J. Opt. Soc. Am. B 3, 125-33 (1986).

[9] U. Keller, Nature 424, 831–838 (2003).

[10] M. Borghesi, A. J. Mackinnon, D. H. Campbell, D. G. Hicks, S. Kar, P.K. Patel, D. Price, L. Romagnani, A. Schiavi, and O. Willi, Phys. Rev. Lett. 92, 055003 (2004).

[11] A. J. Mackinnon, P. K. Patel, M. Borghesi, R. C. Clarke, R. R. Freeman, H. Habara, S. P. Hatchett, D. Hey, D. G. Hicks, S. Kar, M. H. Key, J. A. King, K. Lancaster, D. Neely, A. Nikkro, P. A. Norreys, M. M. Notley, T. W. Phillips, L. Romagnani, R. A. Snavely, R. B. Stephens, and R. P. J. Town Phys. Rev. Lett. 97, 045001 (2006).

[12] T. Sizyuk, A. Hassanein, Phys. Plasmas 19, 083102 (2012).

[13] A. Roy, S.S. Harilal, M.P. Polek, S.M. Hassan, A. Endo, A. Hassanein, Phys. Plasmas 21 033109 (2014).

[14] I. Borgun, N. Azarenkov, A. Hassanein, A. Tseluyko, V. Maslov, D. Ryabchikov, Phys. Lett. A 377, 307 (2013).

[15] H. Daido, M. Nishiuchi, and A. S. Pirozhkov, Rep. Prog. Phys. 75, 056401 (2012).

[16] A. Macchi, M. Borghesi, and M. Passoni, Rev. Mod. Phys. 85, (2013).

[17] E. Fazio, B. Gökce, A. D. Giacomo, M. Meneghetti, G. Compagnini, M. Tommasini, F. Waag, A. Lucotti, C. G. Zanchi, P. M. Ossi, M. Dell'Aglio, L. D'Urso, M. Condorelli, V.

Scardaci, F. Biscaglia, L. Litti, M. Gobbo, G. Gallo, M. Santoro, S. Trusso and F. Neri, Nanomaterials, 10, 2317 (2020).
[18] C. X. Wang and G. W. Yang, Thermodynamic and kinetic approaches of diamond and related nanomaterials formed by laser ablation in liquid. In Laser Ablation in Liquids. Principles and Applications in the Preparation of Nanomaterials. (Jenny Stanford Publishing, New York, 2012).
[19] F. Taccogna, J. Plasma Phys. 81, 495810509 (2015).
[20] A. V. Simakin, V. V. Voronov, and G. A. Shafeev, Phys. Wave Phenom. 4, 218–240 (2007).
[21] N. Haustrup, G. M. O. Connor, Phys. Procedia 12, 46–53 (2011).
[22] N. Tsakiris, K. K. Anoop, G. Ausanio, M. Gill-Comeau, R. Bruzzese, S. Amoruso, and L. J. Lewis, J. Appl. Phys. 115, 243301 (2014).
[23] G. Palazzo, G. Valenza, M. Dell'Aglio, A. D. Giacomo, J. Colloid Interface Sci. 09, 017 (2016).
[24] F. Taccogna1, M. Dell'Aglio1, M. Rutigliano1, G. Valenza and A. De Giacomo, Plasma Sources Sci. Technol. 26, 045002 (2017).
[25] B. Ko, W. Lu, A. V. Sokolov, H. W. H. Lee, M. O. Scully, and Z. Zhang, Sci. Rep. 10, 15753 (2020).
[26] L. Gentile, H. Mateos, A. Mallardi, M. Dell'Aglio, A. D. Giacomo, N. Cioffi, G. Palazzo, J Nanopart Res. 23, 35 (2021).
[27] M. Aliofkhazraei, Handbook of Nanoparticles (Springer, Switzerland, 2016).
[28] M. E. Dieckmann, G. Sarri, L. Romagnani, I. Kourakis, and M. Borghesi, Plasma Phys. Controlled Fusion 52, 025001 (2010).
[29] P. Mora, Phys. Rev. Lett. 90, 185002 (2003).
[30] T. Kiefer, T. Schlegel, and M. C. Kaluza, Phys. Rev. E 87, 043110 (2013).
[31] I. S. Elkamash and I. Kourakis, Phys. Rev. E 94, 053202 (2016).
[32] V. M. Vasyliunas, J. Geophys. Res. 73, 2839 (1968).
[33] M. Hellberg and R. Mace, Phys. Plasmas 9, 1495 (2002).
[34] V. Pierrard and M. Lazar, Sol. Phys. 267, 153 (2010).
[35] G. Sarri, M. E. Dieckmann, C. Brown, C. Cecchetti, D. Hoarty, S. James, R. Jung, I. Kourakis, H. Schamel, O. Willi et al., Phys. Plasmas 17, 010701 (2010).

`

Fig. 1

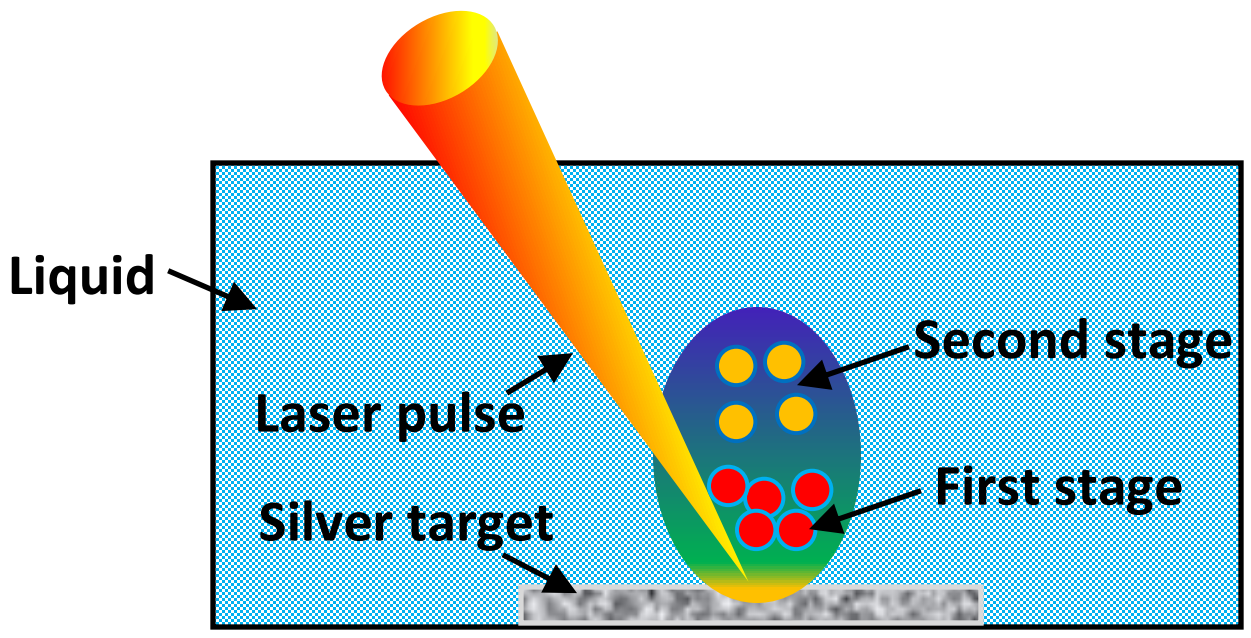

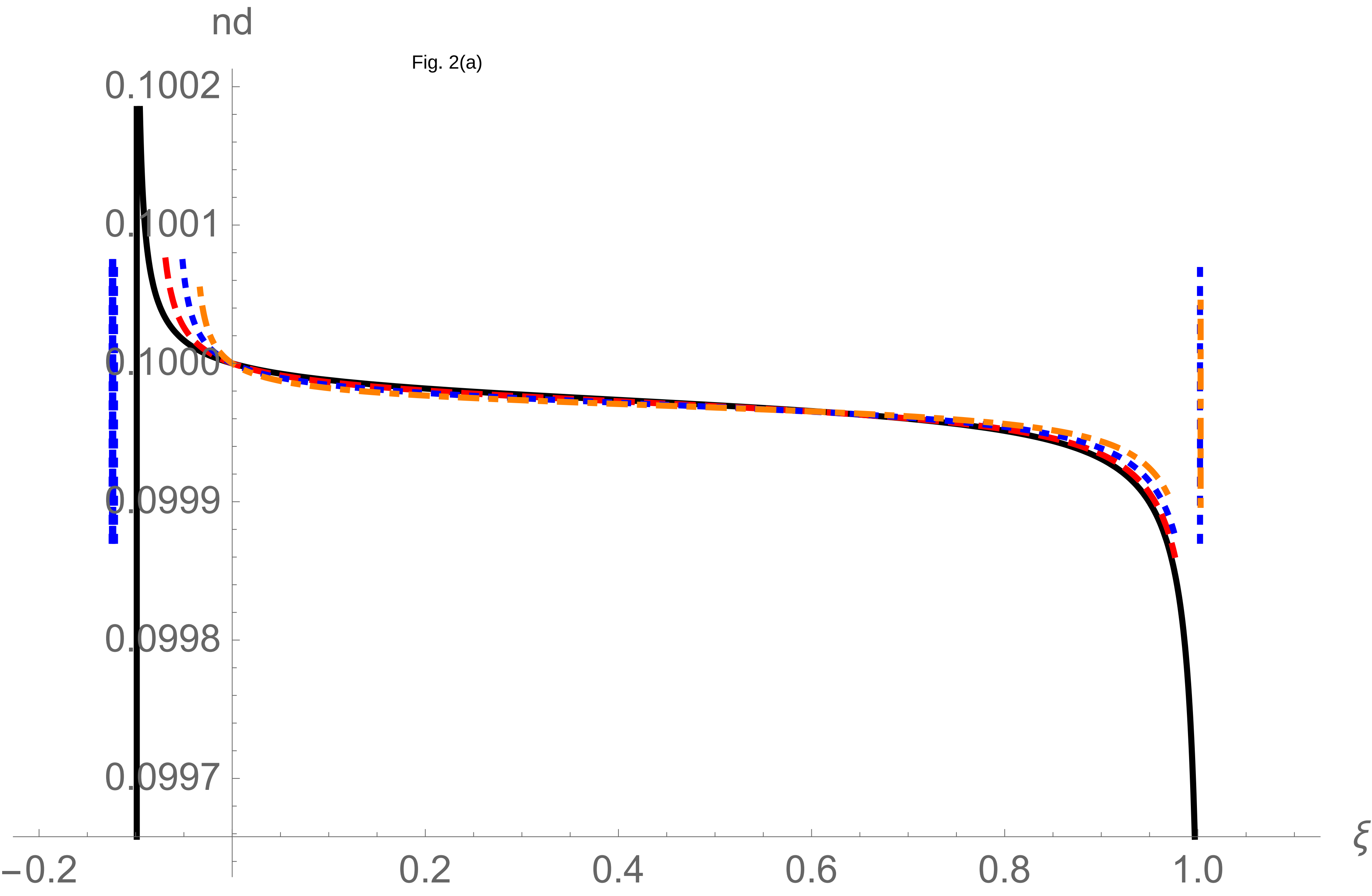

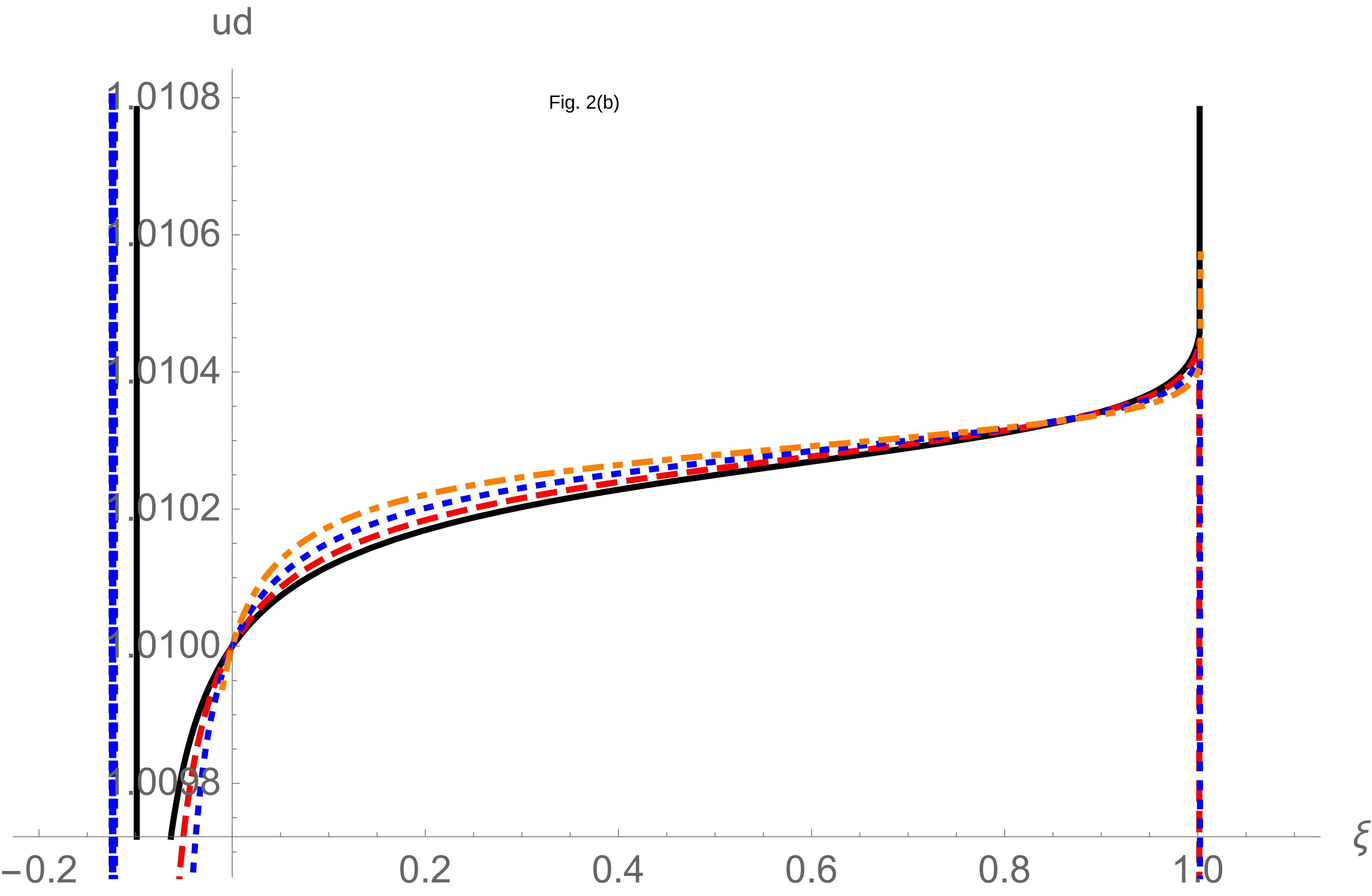

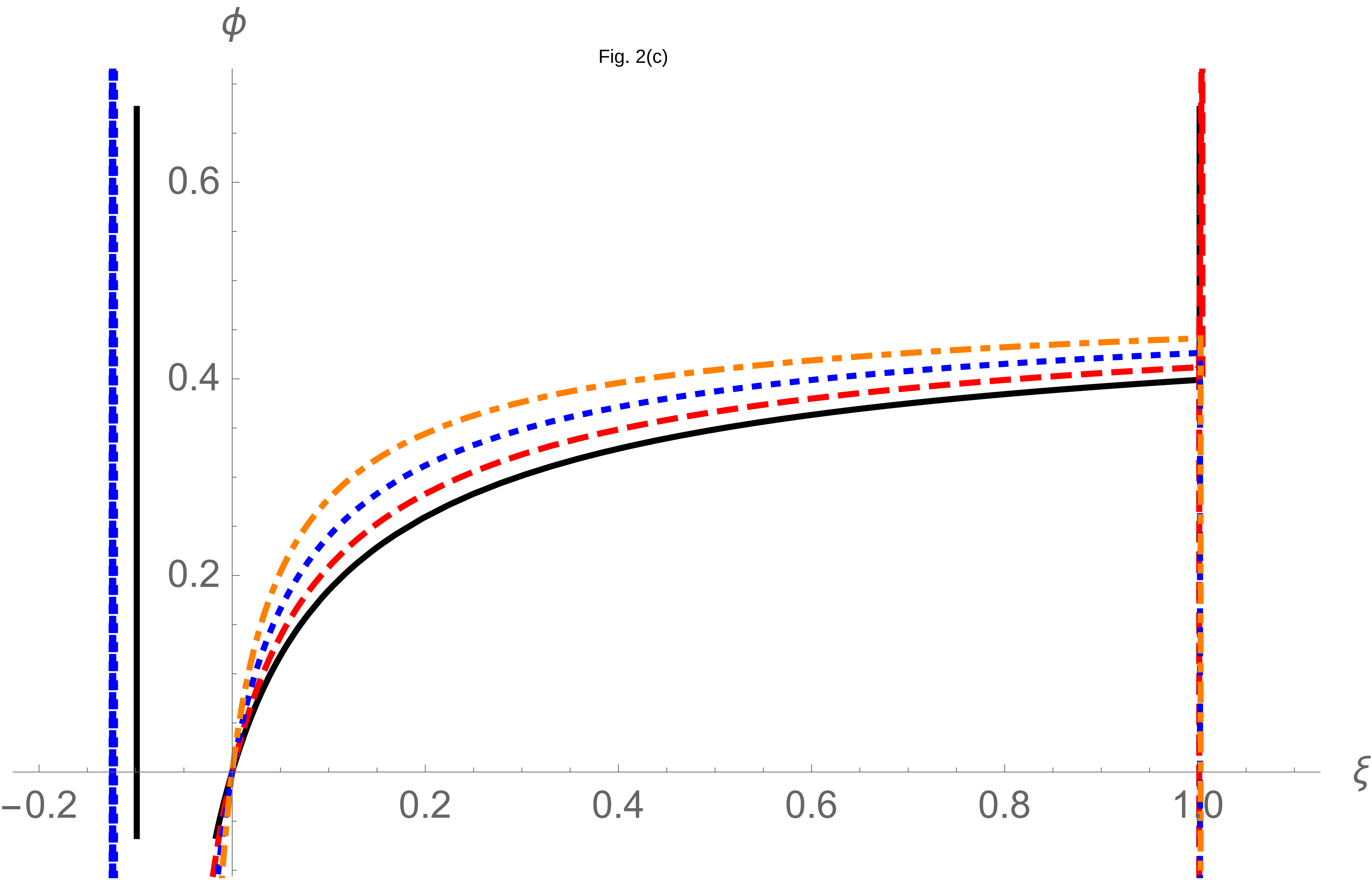

Fig. 2(c)

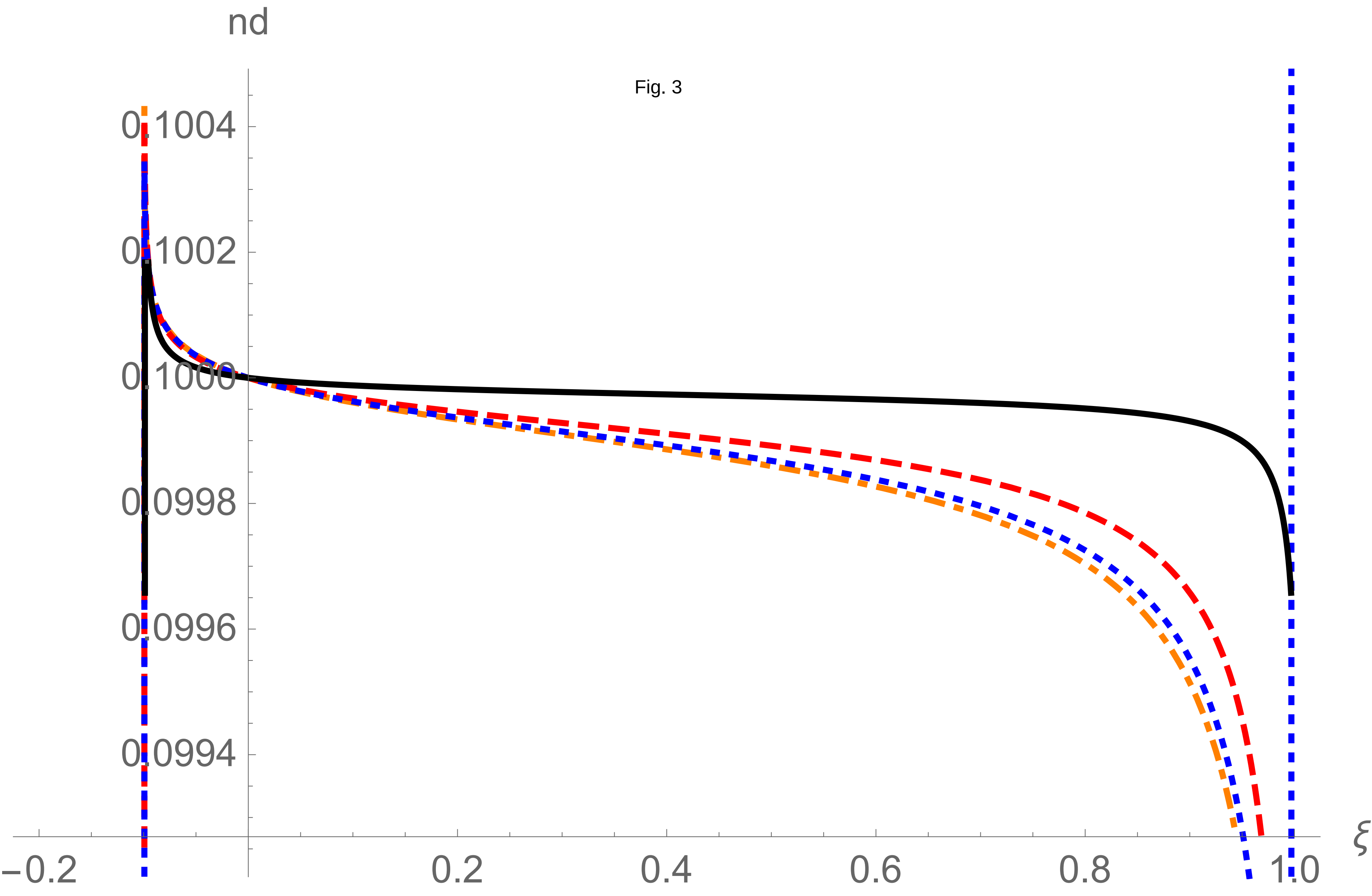

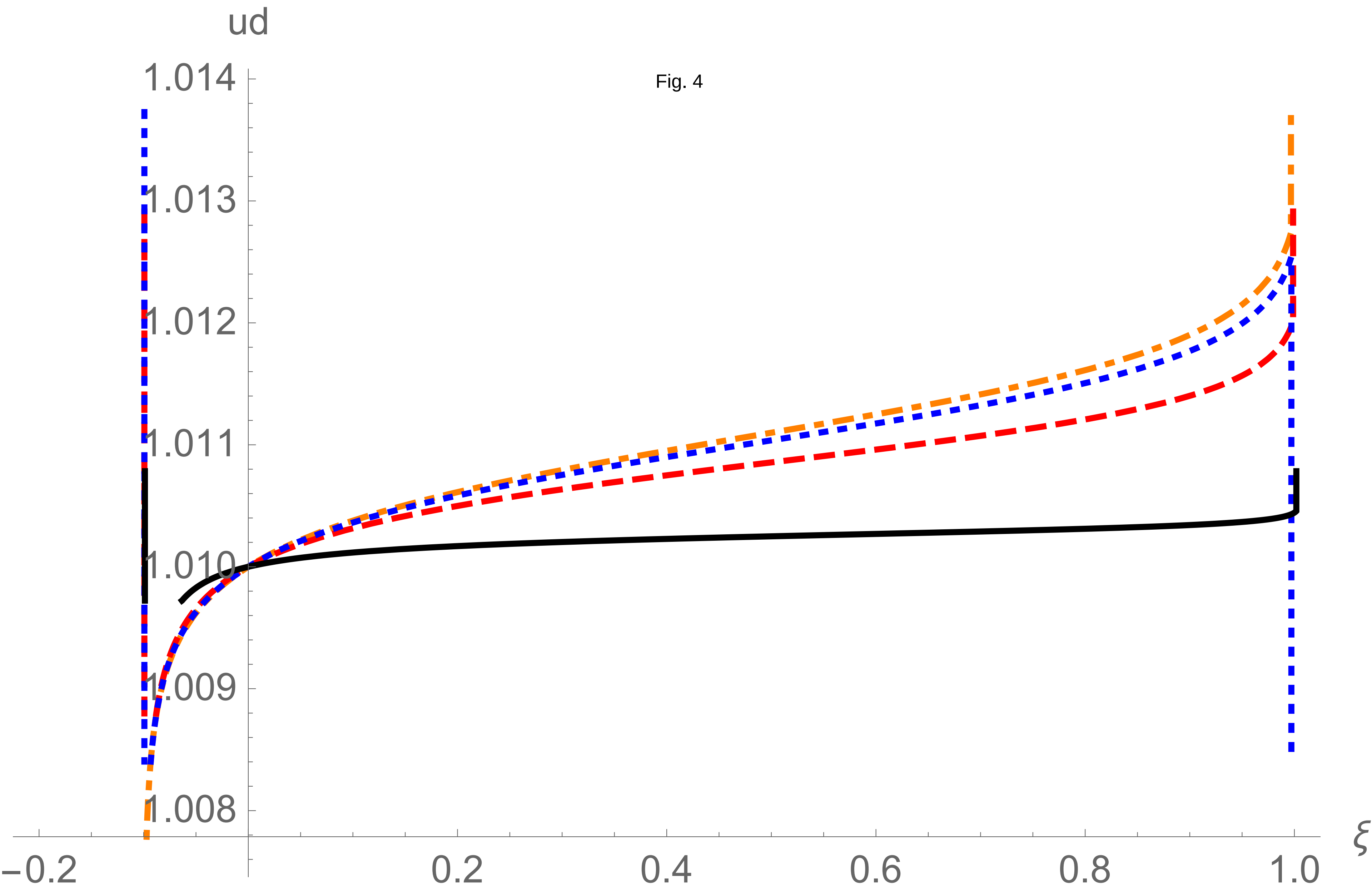

Fig. 5(a)

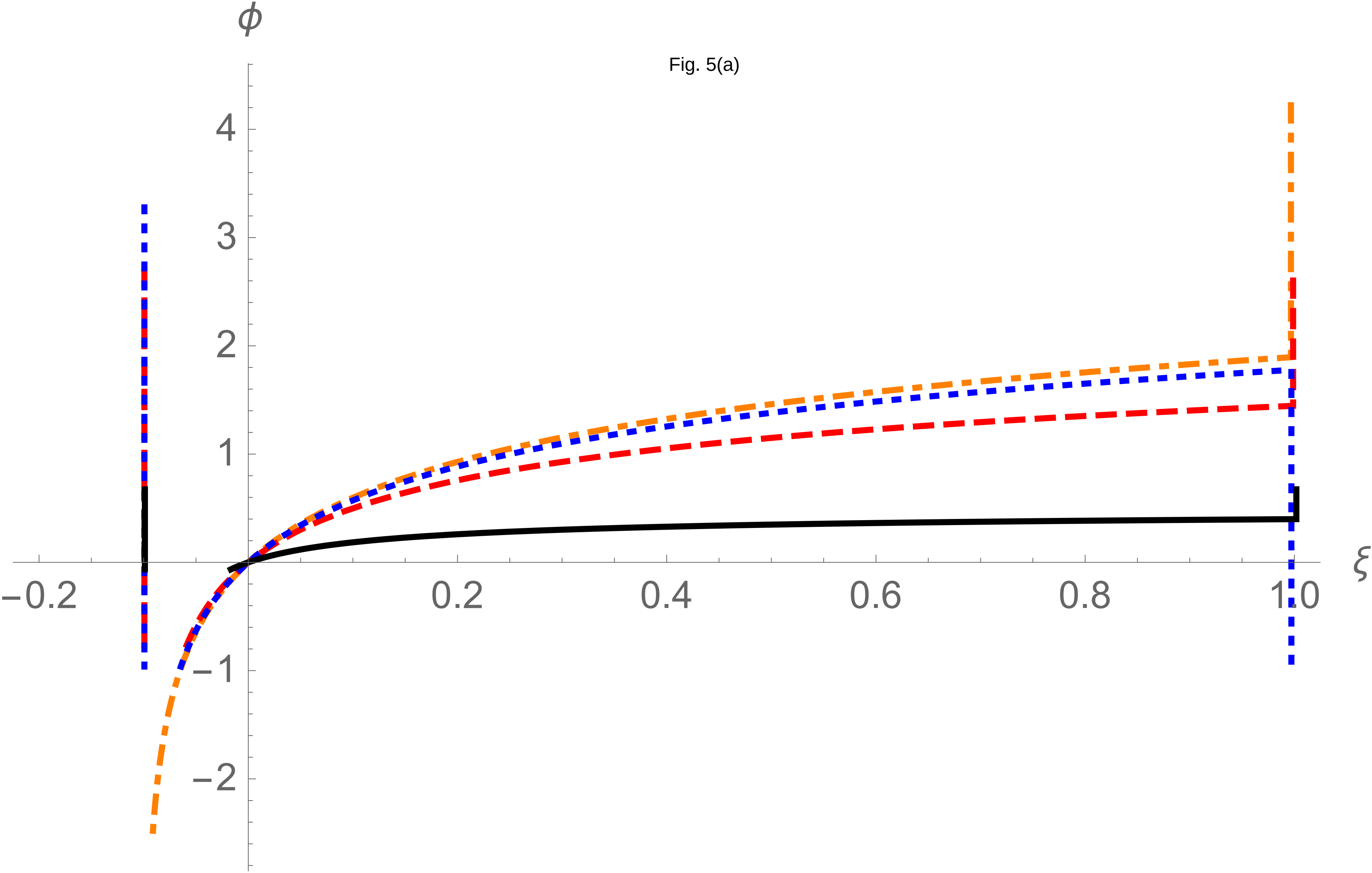

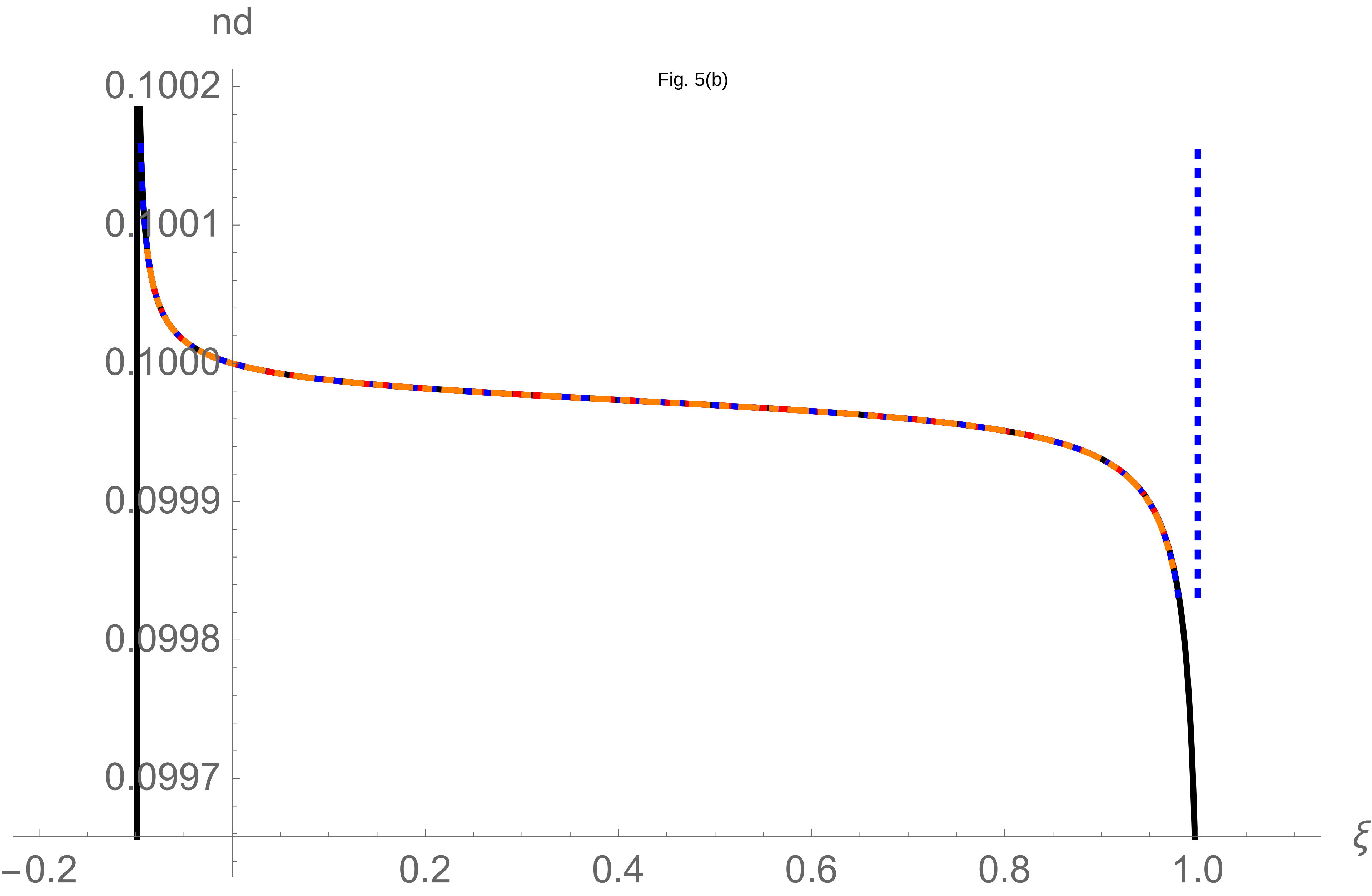

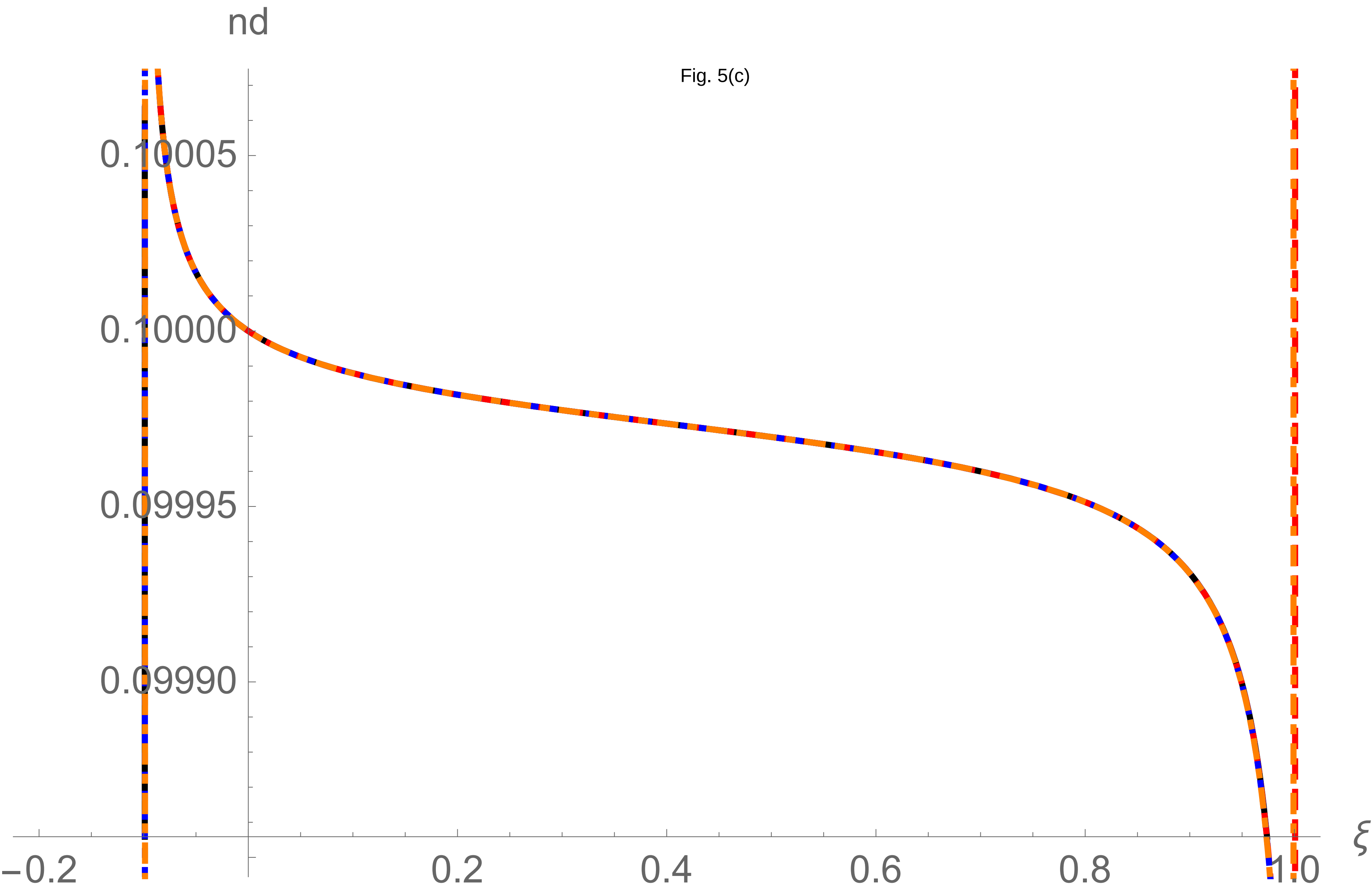

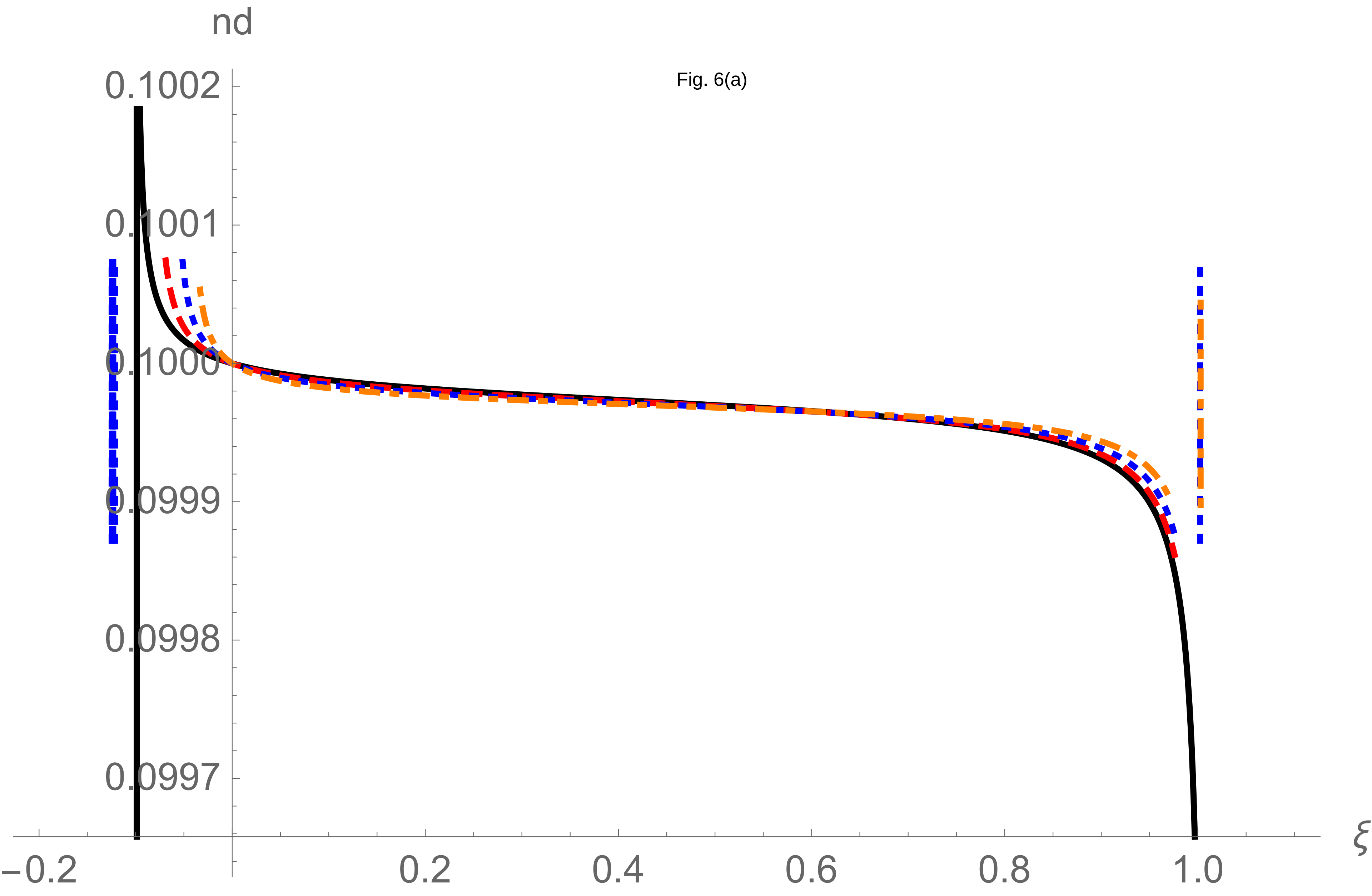

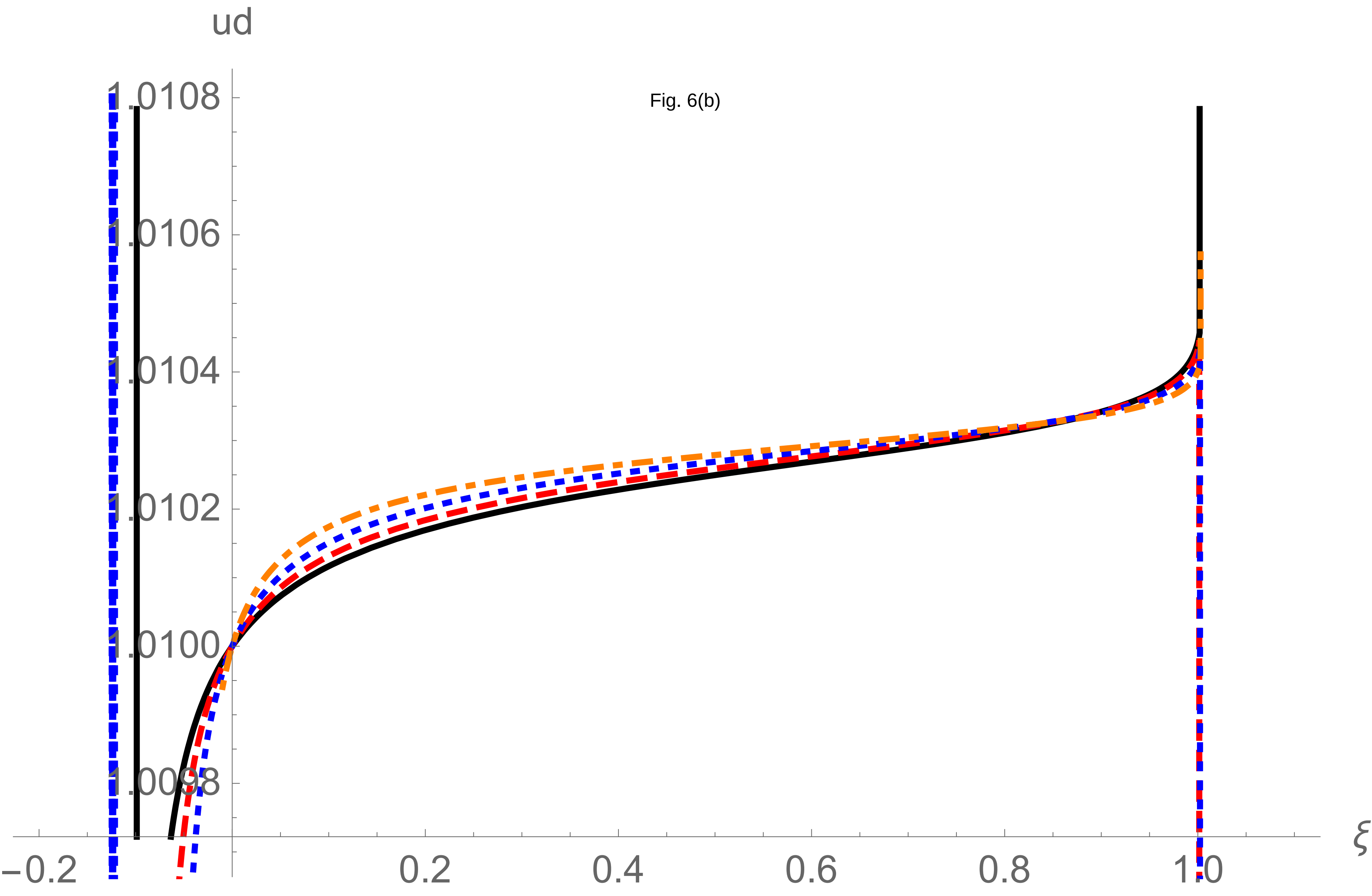

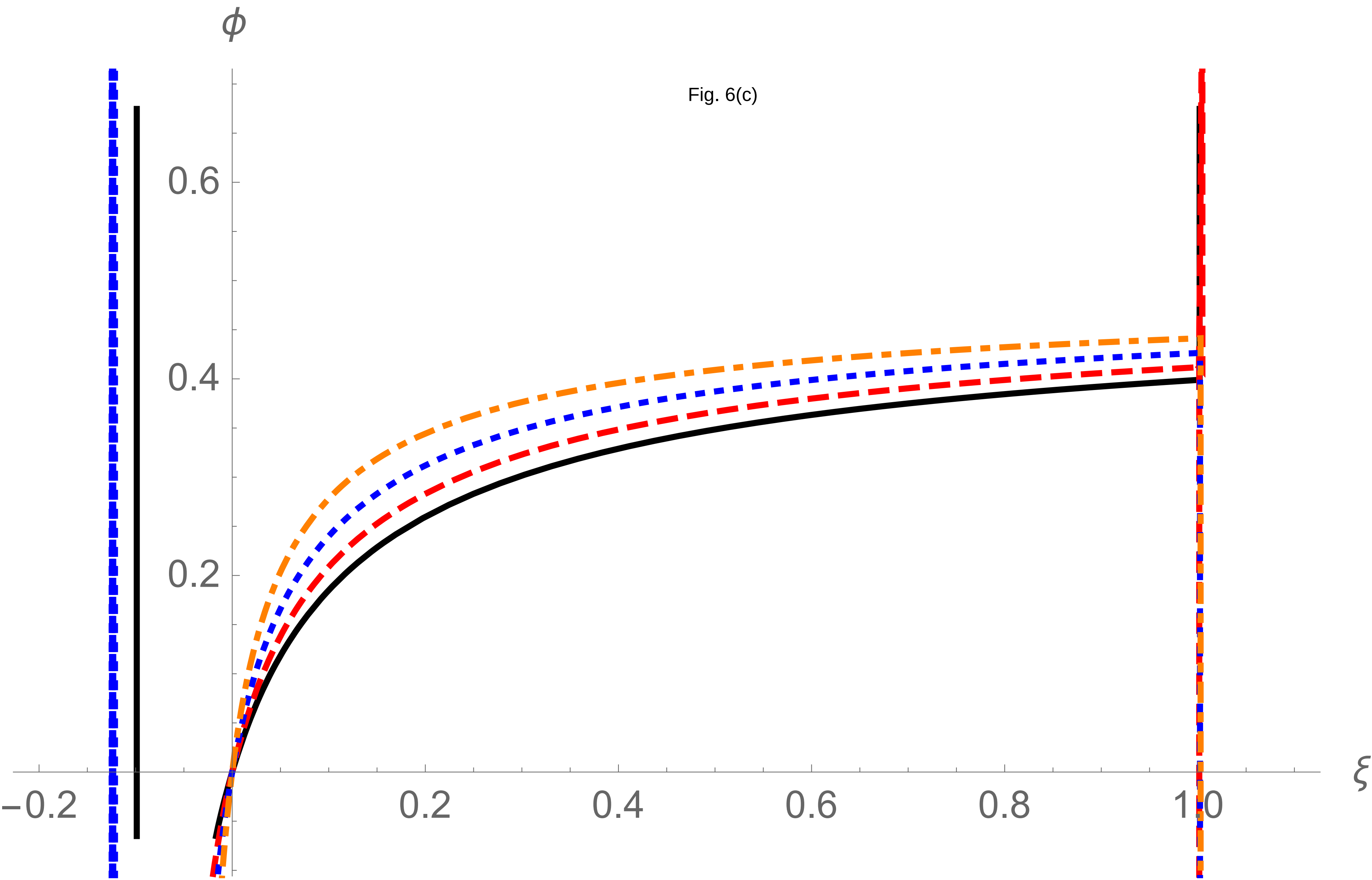

Fig. 7(a)

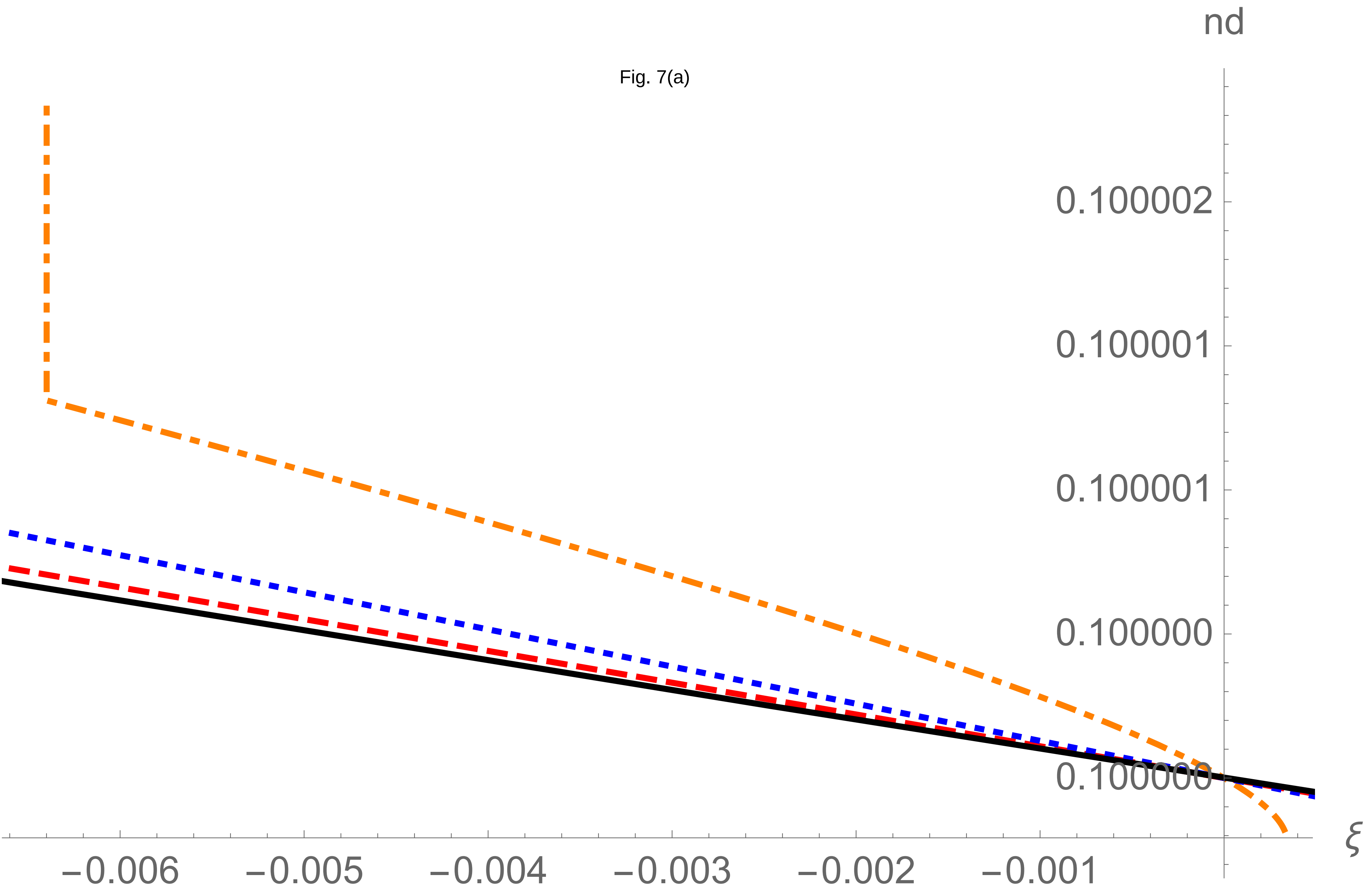

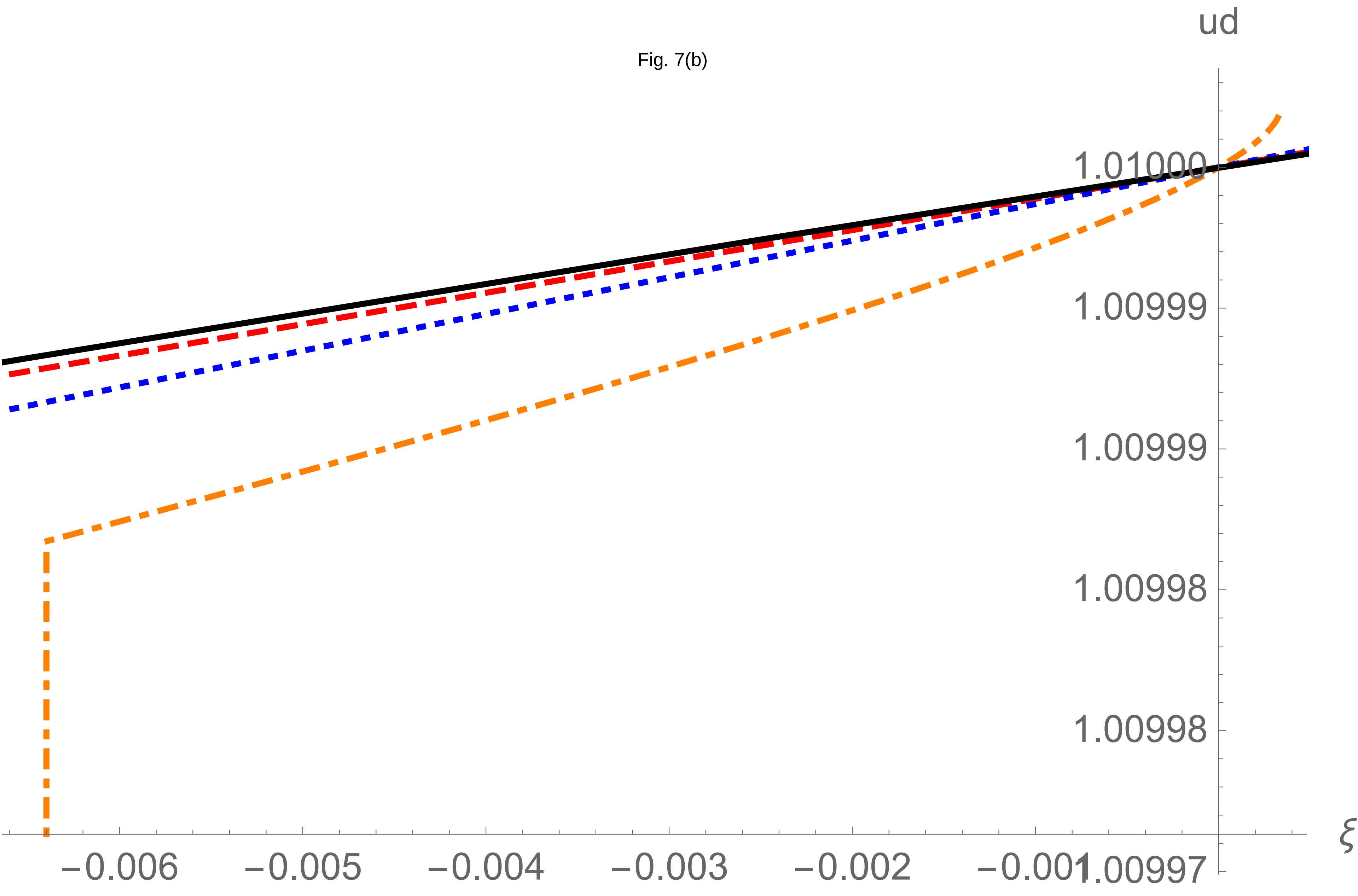

Fig. 7(c)

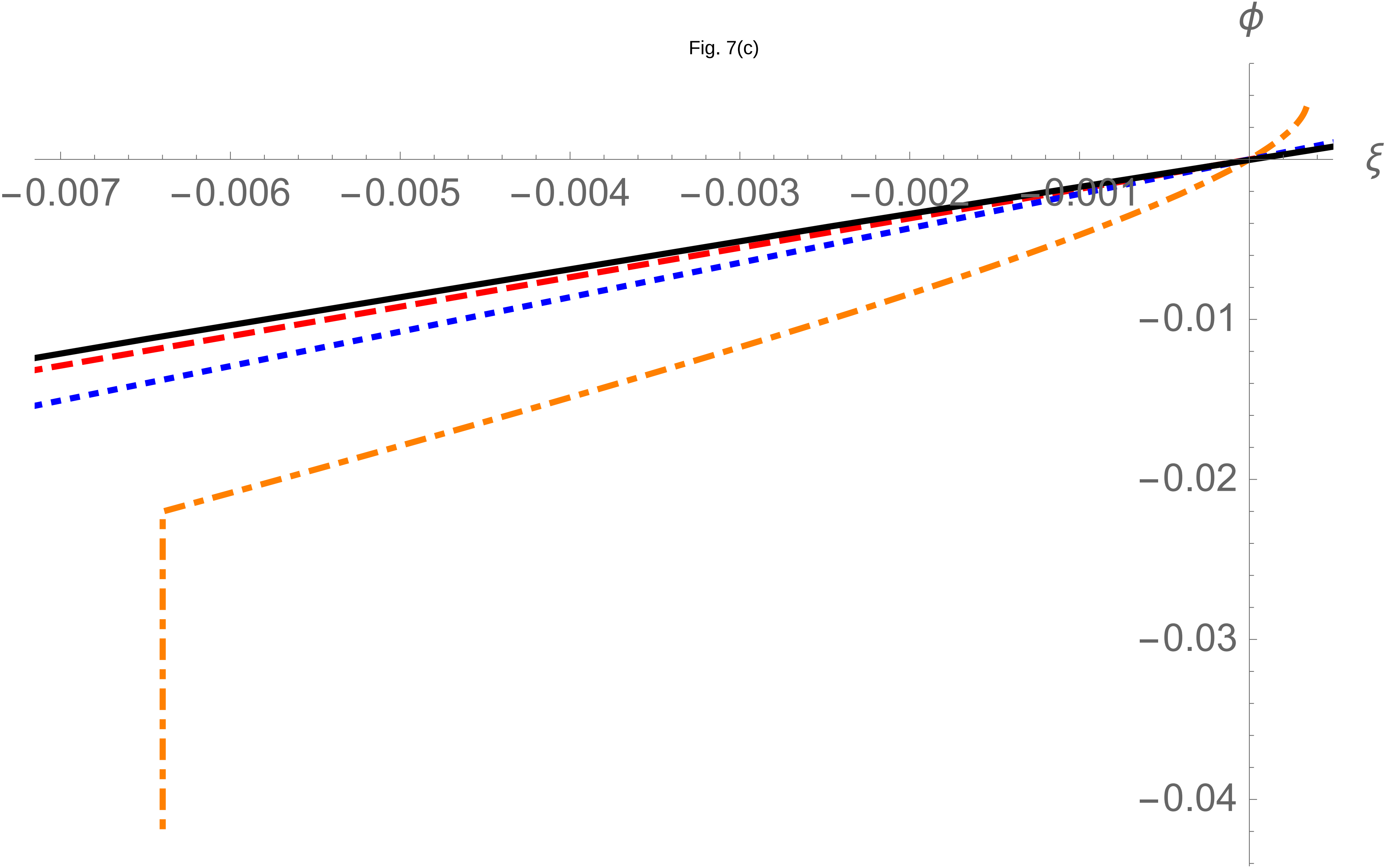

Fig. 8(a)

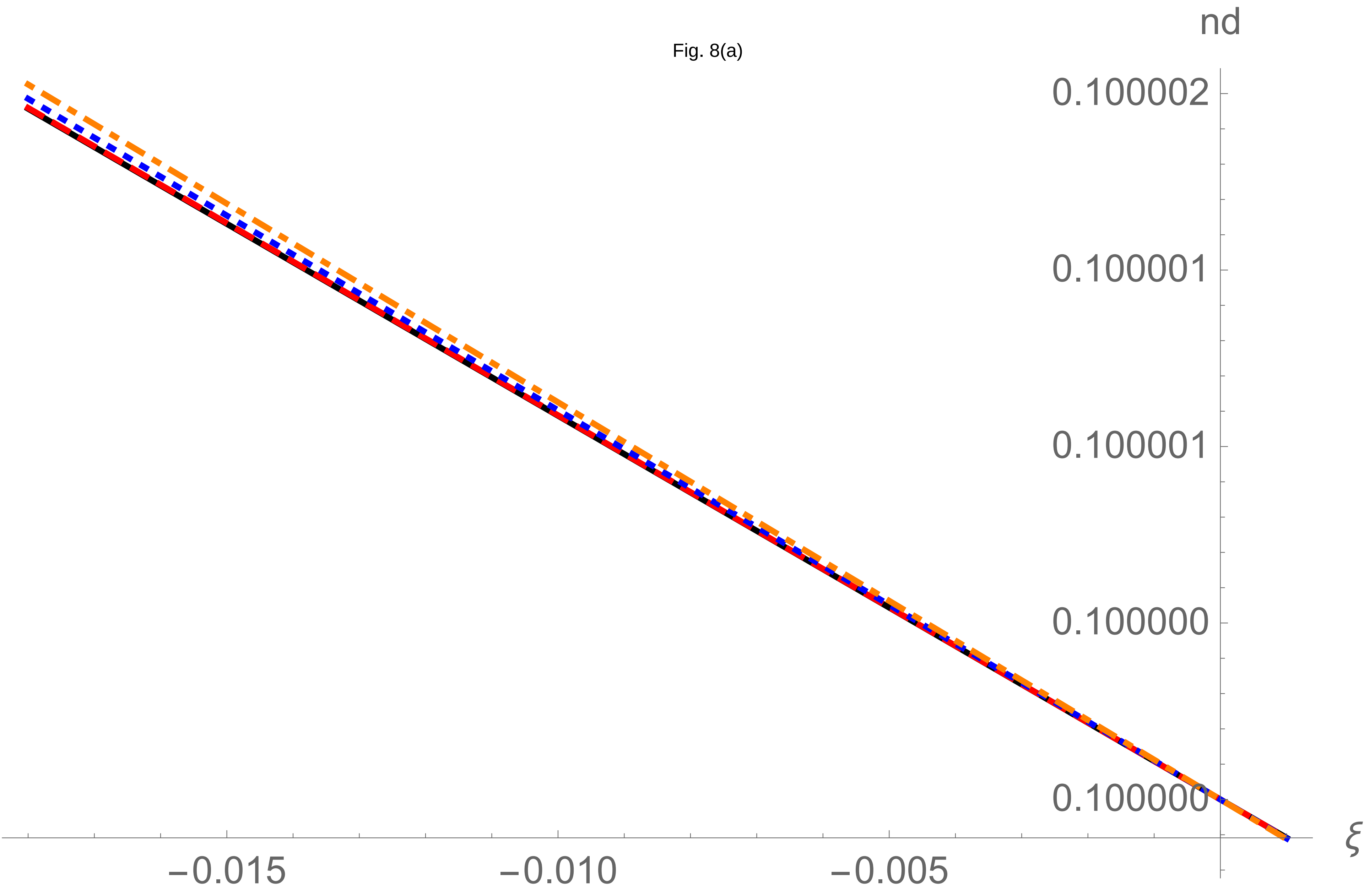

Fig. 8(b)

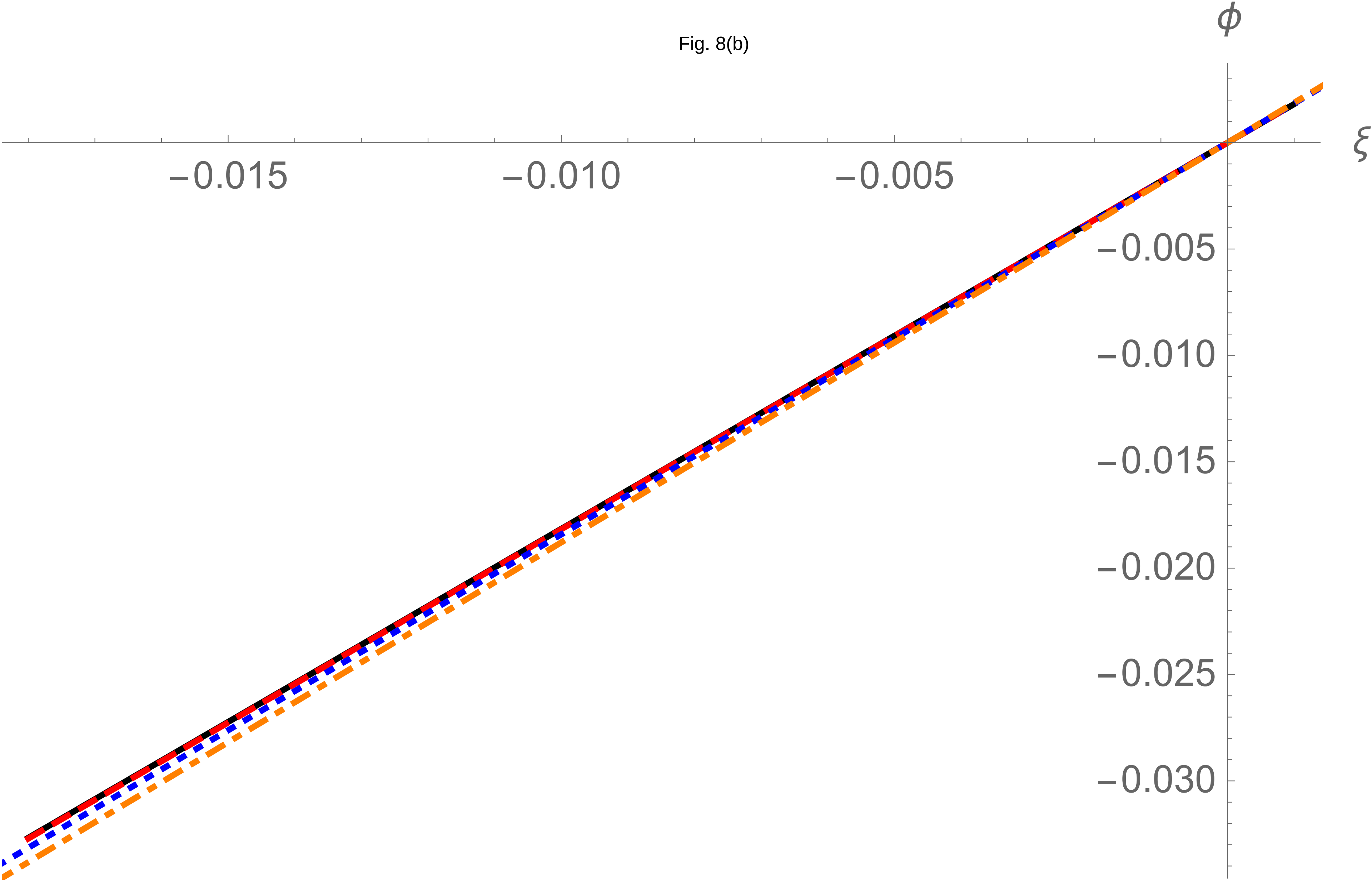

Fig. 8(c)

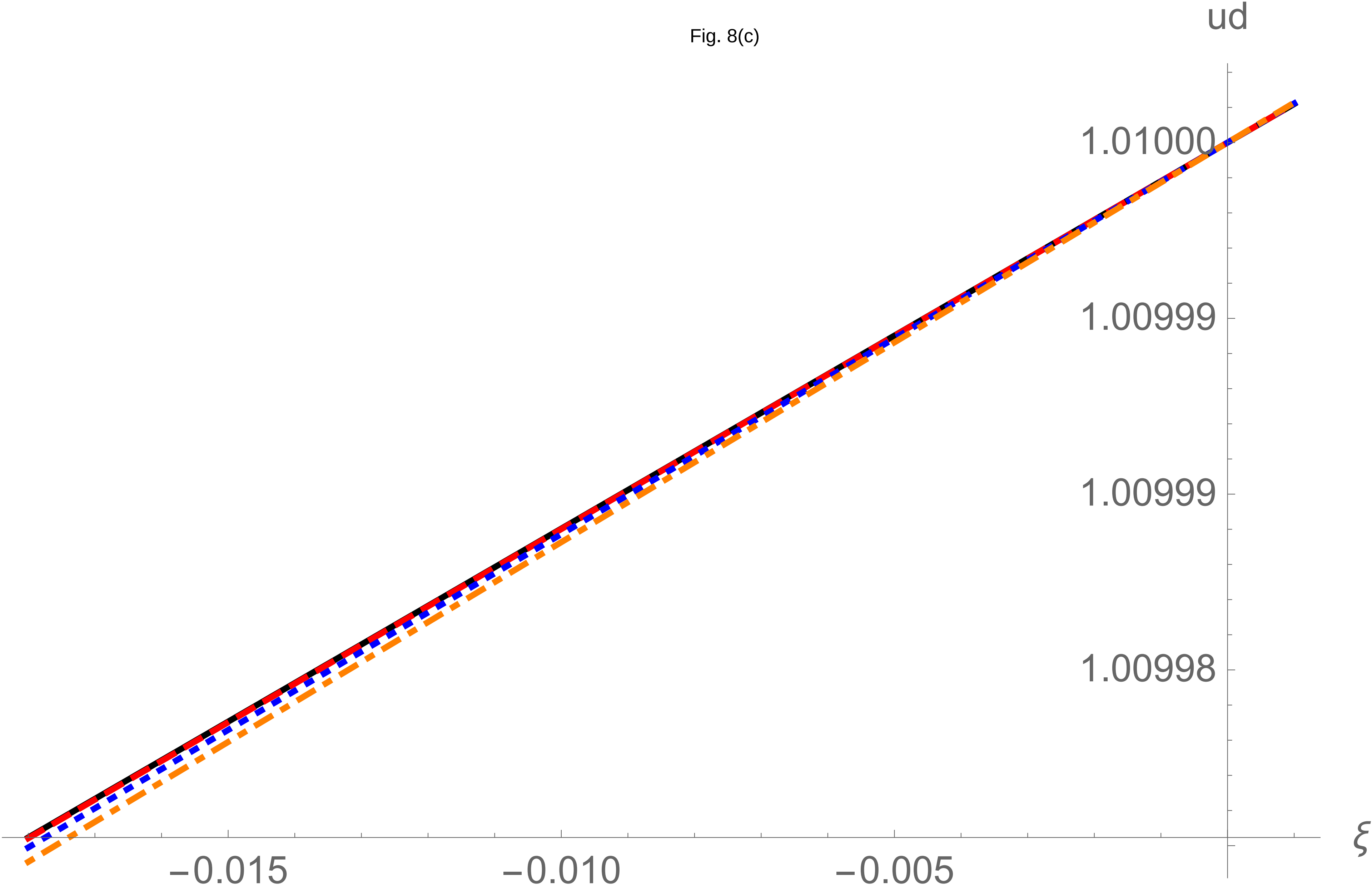

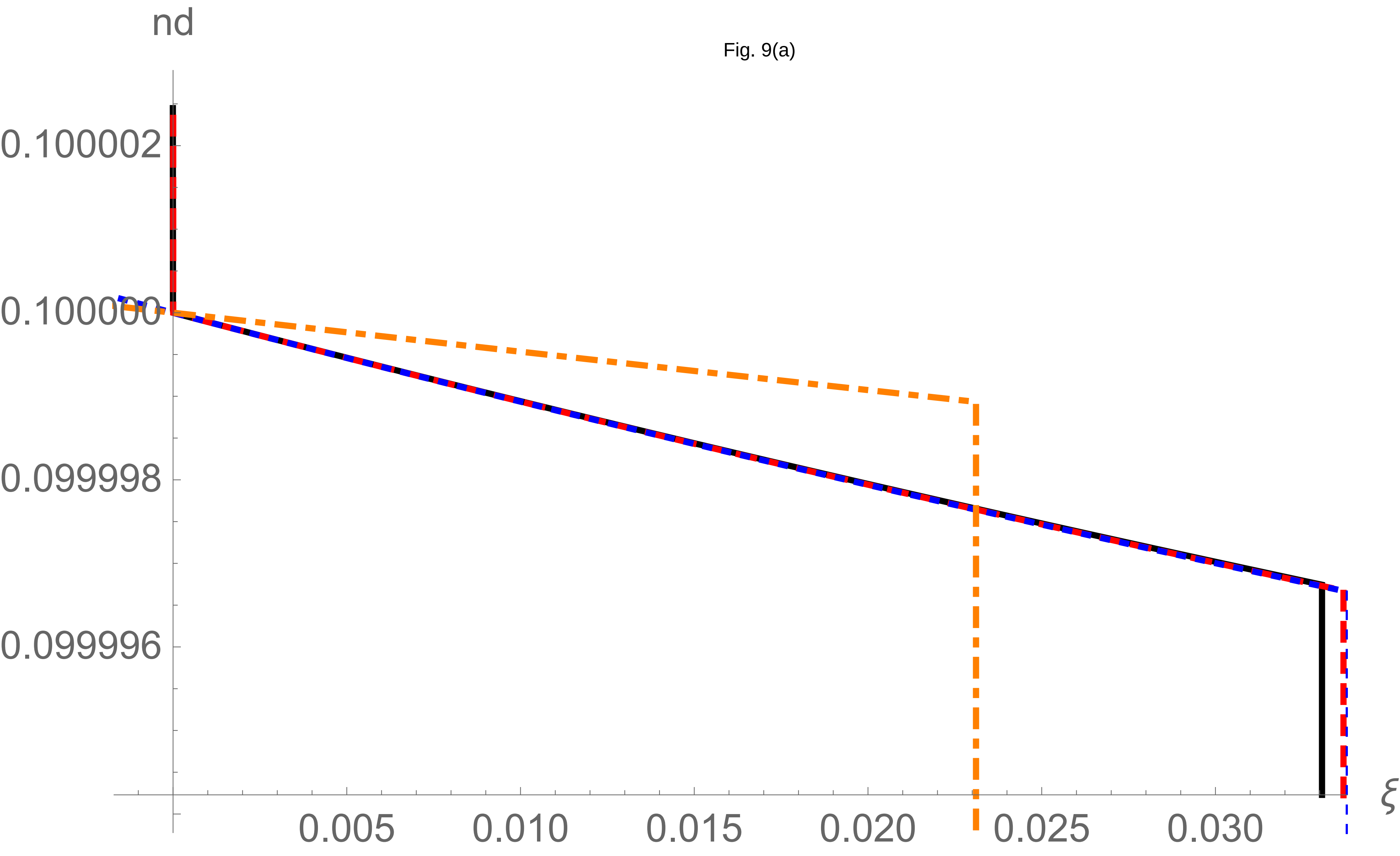

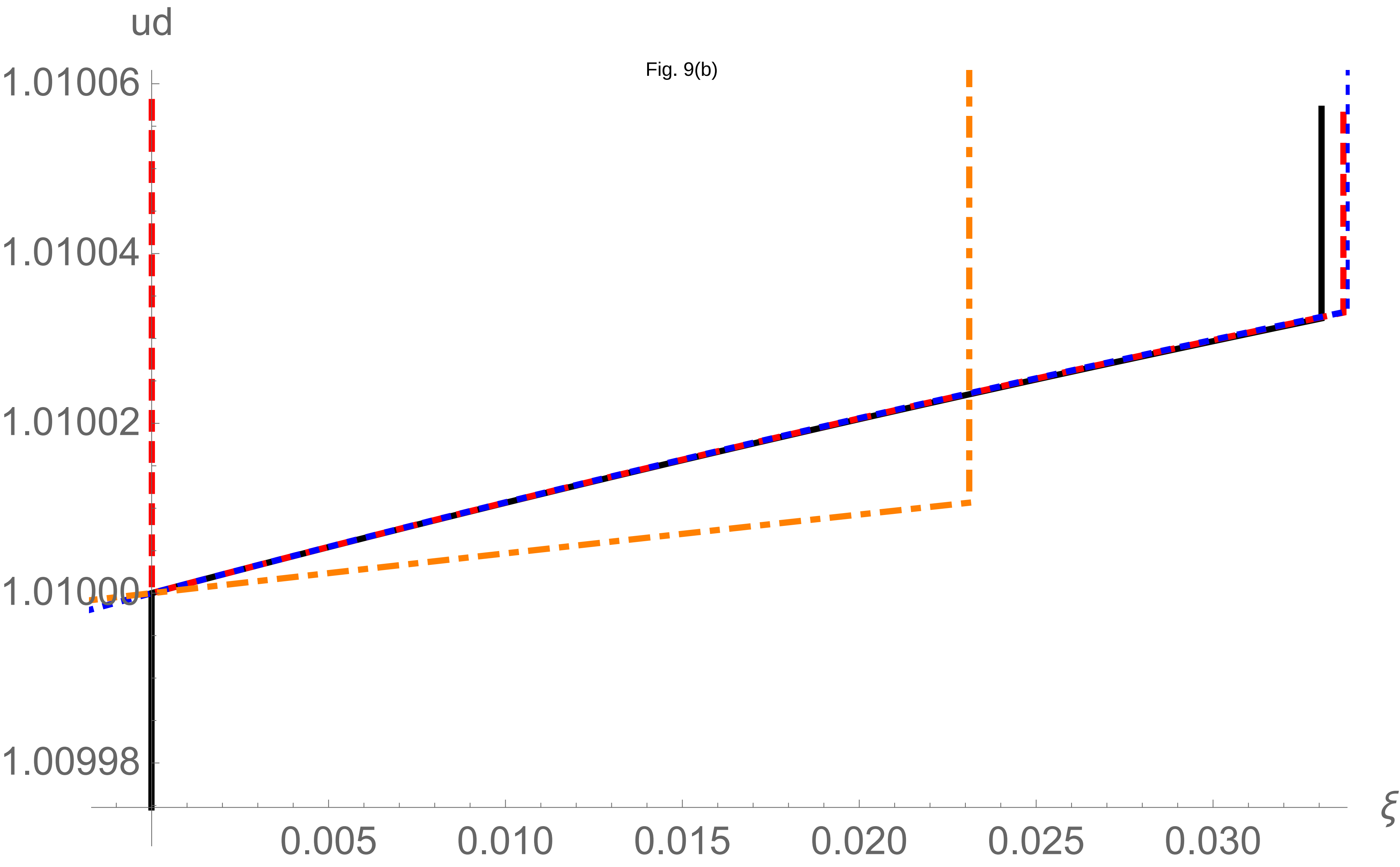

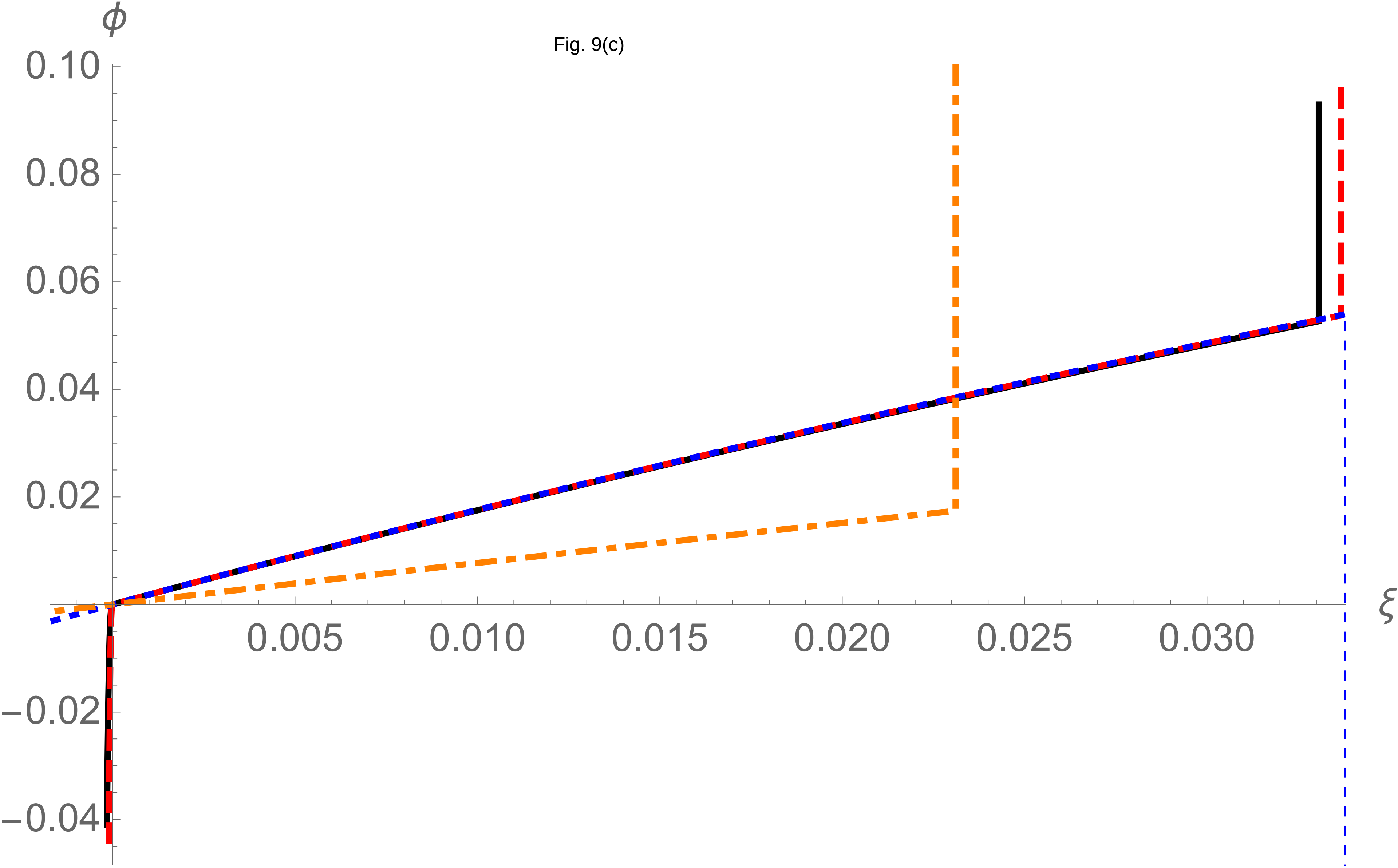

Fig. 9(c)